\documentclass[
a4paper,
twocolumn,longbibliography,nofootinbib,
amsmath,amssymb,
prb,
]{revtex4-1}

\usepackage{graphicx}
\usepackage{dcolumn}
\usepackage{bm}
\usepackage{bbm}
\usepackage{mathrsfs}
\usepackage{hyperref}
\usepackage{csquotes}

\usepackage[table]{xcolor}
\usepackage{ulem} 

\DeclareMathOperator{\Tr}{Tr}

\DeclareMathOperator{\re}{Re}
\DeclareMathOperator{\im}{Im}
\DeclareMathOperator{\rk}{rk}

\DeclareMathOperator{\diag}{diag}

\DeclareMathOperator{\const}{const}

\begin{document}
	
	\title{
		Controlling Quantum Transport in a Superconducting Device via Dissipative Baths}
	
	\author{S.\,V.\, Aksenov$^{1,2}$}
	\email{e-mail: asv@itp.ac.ru}
	\author{M.\,S.\, Shustin$^{1,2}$}
	\author{I.\,S.\, Burmistrov$^{1,3}$}

	\affiliation{%
		$^1$L. D. Landau Institute for Theoretical Physics, 142432 Chernogolovka, Russia\\
		$^2$Kirensky Institute of Physics, Federal Research Center KSC SB RAS, 660036 Krasnoyarsk, Russia\\
		$^3$Laboratory for Condensed Matter Physics, HSE University, 101000 Moscow, Russia}
	
	\date{\today}
	
	\begin{abstract}
		
		Within the quantum field-theoretical approach describing the evolution of a quadratic Liouvillian in the basis of Keldysh contour coherent states, we investigate the spectral and transport properties of a dissipative superconducting system coupled to normal Fermi reservoirs. We derive a generalization of the Meir-Wingreen formula and Onsager matrix for a superconducting system subject to an arbitrary number of fermionic baths. Following Kirchhoff's rule, we obtain an expression describing the dissipation-induced loss current and formulate modified quantum kinetic equations. For wide-band contacts locally coupled to individual sites, we find that each contact reduces the degeneracy multiplicity of the non-equilibrium steady state by one. These results are numerically verified through several cases of the extended Kitaev model at symmetric points with a single contact. Furthermore, in the linear response regime at low temperatures, we demonstrate that (non-)degenerate non-equilibrium steady states correspond to (non-)quantized conductance peaks. Revisiting a 
		paradigmatic problem of resonant transport in the Majorana mode of the Kitaev model we demonstrate that the dissipation accounts for the zero-bias peak suppression and its asymmetry.
	\end{abstract}	
	
	\maketitle
	
	
	\section{\label{sec1}Introduction}
	
	Quantum transport is a universal method for studying the properties of new nanostructures that can shed light on fundamental issues in condensed matter physics and have potential for application in electronics \cite{bercioux-15,arseev-17,jin-23}. 
	Its significance stems from the fact that the observed features of charge and heat currents in a Fermi reservoir (lead) interacting with a sample (device) are directly determined by the spectral characteristics and density of states of the latter \cite{datta-95,haug-96,pekola-21}. The theoretical description of transport phenomena differs depending on the physical processes occurring in the device. The most general approach, capable of accounting for many-body effects and non-linear response, involves examining the time evolution of quantum mechanical operators along the Keldysh contour and calculating current in terms of non-equilibrium Green's functions \cite{keldysh-65,caroli-71a,caroli-71b,combescot-71,caroli-72,meir-92,jauho-94}. 
	
	Although significant progress has been made in developing the non-equilibrium Green's function method since the seminal paper \cite{keldysh-65}, its primary application until recently has been limited to dissipationless systems \cite{arseev-15,kamenev-23b}. 
	In particular, 
	the previous development neglects particle exchange processes between the external environment (henceforth referred to as the ``bath") and the non-equilibrium system comprising leads and device, that makes the total system to be open-like \cite{breuer-02}. 
	
	It is worth noting that 
	phenomenological approaches for describing the effect of incoherent interactions within devices on the quantum transport have been previously proposed. 
	In particular, for modeling phase-breaking electron scattering processes within the Landauer-B?ttiker framework \cite{landauer-57,landauer-70,buttiker-86a}, which describes a specific case of non-equilibrium problems with coherent carrier propagation, additional fictitious Fermi reservoirs (the so-called B?ttiker `probes') were introduced into the transport scheme \cite{buttiker-86b}. These `probes' carry zero net current while exchanging particles with the device.
	
	Since energy loss represents a ubiquitous phenomenon, extending the non-equilibrium Green's function formalism to dissipative processes constitutes an important theoretical challenge. The significance of this task has grown substantially in recent decades due to technological advances in cold atom systems and the consequent expanded possibilities for controlled dissipation engineering \cite{corman-19}. A major obstacle in solving this problem lies in properly accounting for the non-unitary evolution of the density matrix for the combined system (the device and the reservoirs) induced by bath interactions. In general Markovian case, this dynamics is governed by the Gorini-Kossakowski-Sudarshan-Lindblad (GKSL) equation \cite{gorini-76,lindblad-76}. The effects of non-Markovianity on the density matrix dynamics, which 
		can be relevant for experiments, are also actively discussed \cite{rivas-14,breuer-16,devega-17,zhang-25b}. 
	
	The derivation of the Keldysh partition function for open bosonic systems \cite{sieberer-14,sieberer-16} 
	demonstrated that the action of a system whose density matrix satisfies the GKSL equation contains not only the conventional Hamiltonian but also a dissipator determined by Lindblad fields that induces non-unitary evolution. This finding triggered developments in transport theory for non-superconducting devices with decay and decoherence inside \cite{visuri-22,visuri-22b,uchino-22,sur-25}, as well as for dissipative superconducting quantum point contacts \cite{visuri-23}. Another active research direction in the non-equilibrium physics, driven by advances in quantum optics, concerns quantum transport theory with Markovian-type leads whose evolution timescales are much shorter than those of the device \cite{bertini-21,kolovsky-21,landi-22,maksimov-23}. Recent results in this area have also been obtained using quantum field-theoretical approach \cite{jin-20}.
	
	Some studies of dissipative superconducting quantum point contacts and Josephson junctions have appeared recently \cite{visuri-23,li-25}.
	Despite these advances, to the authors' knowledge, there is currently no comprehensive theory of quantum transport in superconducting devices with dissipation. Nevertheless, the field of superconducting nanostructures has experienced remarkable growth in recent decades \cite{linder-15,nadeem-23,maggiora-24,amundsen-24}. This development has been largely driven by efforts to establish scalable quantum information processing schemes \cite{wendin-17}, including those based on novel principles of topological qubit protection \cite{kitaev-03,nayak-08}.
	
	Hybrid semiconducting/superconducting nanowires have emerged as one of the most promising platforms for realizing topological quantum computing \cite{alicea-11,plugge-17} and related search of Majorana states \cite{lutchyn-10,oreg-10,mourik-12,aghaee-23}. However, despite numerous experimental studies \cite{mourik-12,aghaee-23}, conclusive evidence for Majorana-type subgap quasiparticles remains elusive \cite{dassarma-23,frolov-23,aksenov-24}.
	
	One potential explanation for this challenge may lie in the presence of fermionic baths interacting with the superconducting wire. In practice, several dissipation mechanisms can be identified:
	(a) Scattered radiation, cosmic rays, and disorder in the parent superconductor 
	responsible for above- and subgap states at the semiconductor/superconductor interface. These states may be interpreted as a fermionic bath if weak interaction and/or finite temperature is present \cite{rainis-12,liu-17}.
	(b) Disorder in the semiconducting core can produce multiple subgap states within the bulk. Transported particles undergoing Andreev reflection at the wire's edge states may scatter into these bulk excitations through different interaction mechanisms \cite{dassarma-16,cole-16}. Relevant interactions include phonon scattering and charge noise from gate electrodes, which become particularly significant when the hybridization of Majorana mode wavefunctions is non-negligible \cite{knapp-18,aseev-18,karzig-21}.
	(c) Subgap states may emerge in vortices in the parent superconductor in the presence of strong enough magnetic field \cite{liu-17,dassarma-16}.
	
	When analyzing spectral and transport properties of superconducting wires, the influence of fermionic baths is often phenomenologically modeled through the introduction of a Dynes parameter in the Hamiltonian \cite{dassarma-16,dynes-78,day-25}. In this work, we develop a microscopic quantum field theory of charge and energy transport in superconducting system (or device) subject to dissipative processes (see Fig. \ref{fig1}). Our results exhibit considerable universality for several reasons and are applicable to a broad class of non-equilibrium systems described by quadratic Liouvillians. First, our treatment accounts for 
	hybridization between the device and an arbitrary number of Fermi reservoirs (i.e. leads) with various 
	configurations. Second, we consider coupling to a 
	finite number of fermionic baths with spatially varying particle gain/loss amplitudes. Third, although for concreteness the device region is described by an extended Kitaev model \cite{kitaev-01,thompson-23} incorporating both normal and anomalous interactions between arbitrary neighbors, our methodology can be applied to arbitrary fermionic system with superconducting correlations.
	
	The paper is organized as follows. In Sec. \ref{sec2} the model of considered superconducting dissipative system is formulated. We present the main results 
	for transport 
	in superconducting dissipative system in Sec. \ref{sec3}. The effect of wide-band leads on the zero-energy excitations in the system is considered in Sec. \ref{sec4}. In Sec. \ref{sec5} 
	our general theory is applied to study the transport features of topological superconductors affected by dissipation. Section \ref{sec6} is devoted to discussions and conclusions. Appendix \ref{appA} is focused on detailed derivation of currents in normal lead connected to the superconductor that exchanges carriers with other leads and fermionic baths as well. In Appendix \ref{appB} we analyze the conditions for non-decaying excitations in the case of a few leads and baths coupled to the device.

	\section{Model}\label{sec2}

\begin{figure}[t]
		\begin{center}
			\includegraphics[width=0.49\textwidth]{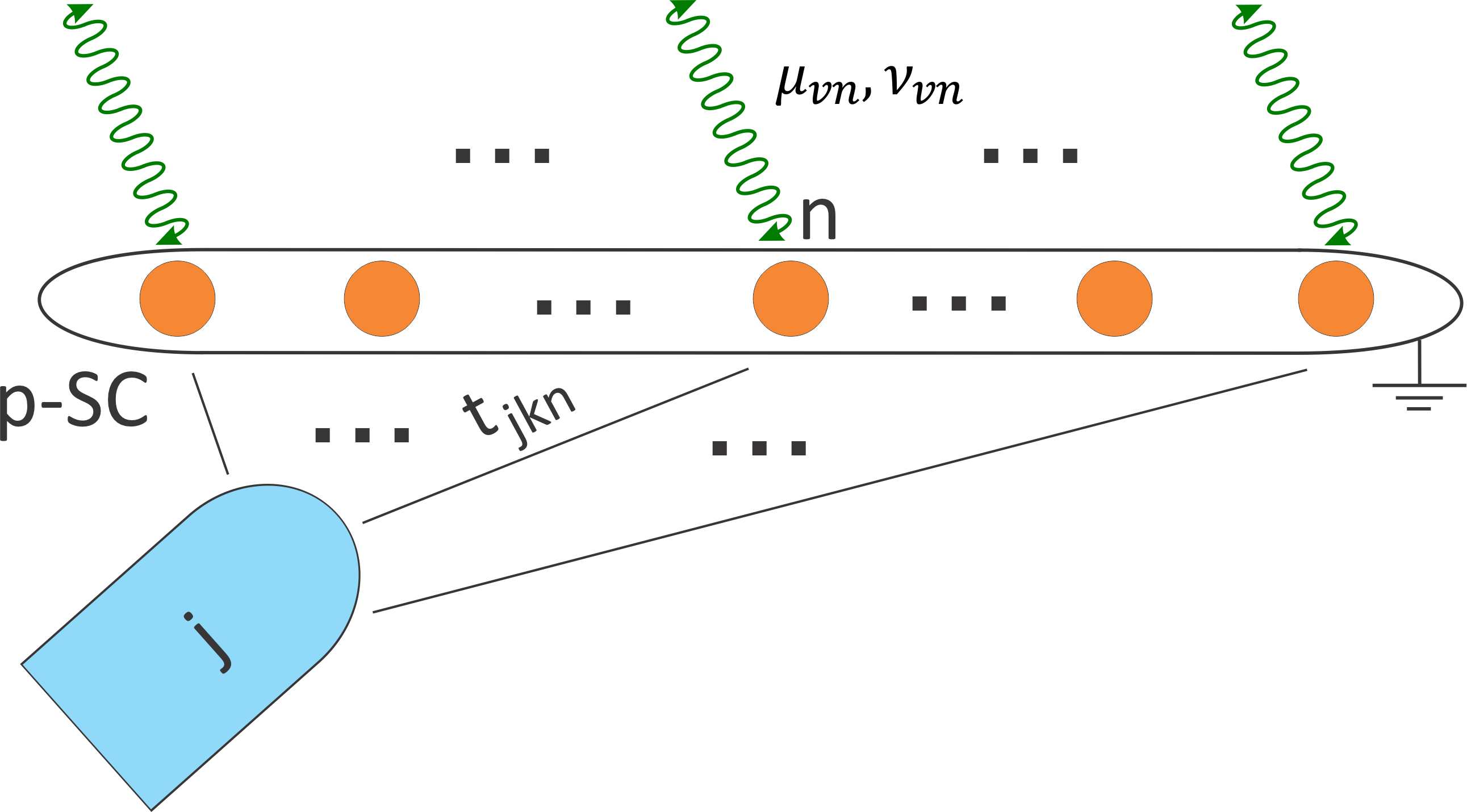}
			\caption{\label{fig1} Sketch of the model. 
				A $p$-wave superconducting device interacts with arbitrary number of fermionic reservoirs (or leads) and baths. Transport problem is analyzed in general situation when each lead and bath are coupled with all chain sites. The interaction parameters at the $n$-th site are $t_{jkn}$ and $\mu_{vn},\nu_{vn}$ for the $j$-th lead and $v$-th bath, respectively.}
		\end{center}
	\end{figure}

	Let us consider a superconducting system of spinless fermions (device) coupled to an arbitrary number of single-band normal leads and baths. The sketch of the system is shown in Figure \ref{fig1}. We suppose the device is described by quadratic tight-binding Hamiltonian of the most general form \cite{bogoliubov-07,thompson-23}:
	\begin{equation} \label{Hs}
		\hat{H}_{s} =\sum\limits_{n,m=1}^{N}\left(\xi_{nm}\hat{c}^+_{n}\hat{c}_{m}+\Delta_{nm}\hat{c}_{n}\hat{c}_{m}+\Delta_{nm}^{*}\hat{c}_{m}^{+}\hat{c}_{n}^{+}\right),
	\end{equation}
	where $\xi_{n=m}=\varepsilon_n-\mu$ stands for an electron energy at the $n$-th site, $\xi_{n\neq m}$ is a hopping 
	amplitude between sites $n$ and $m$, $\Delta_{nm}$ denotes a superconducting pairing parameter for electrons located at the sites $n$ and $m$. The Bogoliubov-de Gennes matrix of the device Hamiltonian is written as: 
	\begin{equation} \label{HsBdG}
		H_{s} =\frac{1}{2}\left( \begin{array}{*{20}{c}}
			\xi & ~~~\Delta^{+} \\
			\Delta & -\xi^{T} \\
		\end{array} \right)\equiv \frac{1}{2} h_{s},
	\end{equation}
	where the diagonal and off-diagonal blocks obey $\xi=\xi^{+}$, $\Delta=-\Delta^{T}$, respectively.
	
	We assume non-interacting leads that are described by the following diagonal Hamiltonians
	\begin{equation} \label{HC}
		\hat{H}_{j} =\sum\limits_{k}\xi_{jk}\hat{d}^+_{jk}\hat{d}_{jk}\equiv\sum\limits_{k}H_{jk},
	\end{equation}
	where $\hat{d}_{jk}$ denotes an electron annihilation operator 
	for a state labeled by $k$ and with the energy $\xi_{jk}=\varepsilon_{jk}-\mu_{j}$ in the $j$-th lead, $\mu_{j}=\mu + eV_{j}$ is an electrochemical potential in $j$-th lead, 
	$V_{j}$ is a bias voltage  applied to the  $j$-th lead (it is assumed that the superconducting system is directly grounded), and $\mu$ is the chemical potential.

	The interaction between the dissipative system and 
	the $j$-th lead is described by the following Hamiltonian
	\begin{equation} \label{HT}
		\hat{H}_{t,j} =\sum\limits_{n=1}^{N}\sum\limits_{k}t_{jkn}\hat{d}_{jk}^{+}\hat{c}_{n}+ {\rm h.c.},
	\end{equation}
	where $t_{jkn}$ is 
	an amplitude characterizing quasiparticle's tunneling between the $j$-th lead and the $n$-th site of the superconducting system. Thus, the Hamiltonian of the system coupled to $N_{L}$ leads is given by 
	\begin{equation} \label{Htot}
		\hat{H}=\hat{H}_{s}+\sum\limits_{j=1}^{N_{L}}\left(\hat{H}_{j}+\hat{H}_{t,j}\right).
	\end{equation}
	
	The presence of dissipative processes in the system leads to a situation where the total density matrix of the system and leads, $\hat{\rho}$, no longer satisfies the von Neumann equation, but rather obeys the GKSL equation (or Lindblad master equation),
	\begin{equation}\label{Lindblad_eq}
		\frac{d\,\hat{\rho}}{d\,t} = -i\,[\,\hat{H},\,\hat{\rho}\,] + \sum_{v=1}^{N_B}\left(\,\hat{L}_{v}\hat{\rho} \hat{L}^+_{v} - \frac{1}{2}\{\hat{L}_{v}^+\hat{L}_{v},\hat{\rho}\}\right),
	\end{equation}
	where the second term in the right side of \eqref{Lindblad_eq} represents the dissipator responsible for non-unitary dynamics. Its properties are determined by the jump operators (or Lindblad operators). Since our goal is to construct a widely applicable quadratic dissipative-transport theory for superconducting systems, we focus on 
		the most general form of the linear dependence of $\hat{L}_{v}$ on $\hat{c}_{n}$ and $\hat{c}_{n}^{+}$ \cite{thompson-23,shustin-25}. In other words, we consider
	
	
	\begin{eqnarray}\label{Lvn}
		\hat{L}_{v}=\sum\limits_{n}\left(\mu_{vn}\hat{c}_{n}+\nu_{vn}\hat{c}_{n}^{+}\right),
	\end{eqnarray}
	where $\mu_{vn},~\nu_{vn} \in \mathbb{C}$ - amplitudes of electron loss and gain at the $n$-th site due to interaction with $v$-th bath. Unless stated otherwise, in what follows we will assume that these amplitudes vary randomly from site to site. In numerical calculations 
		the real and imaginary parts of $\mu_{vn}$ and $\nu_{nv}$ obey a standard uniform distribution (see the details in Sec.\ref{sec5}).
	Note the right side of the GKSL equation can be formally treated as an action of a Liouvillian superoperator on the density matrix, $d\hat{\rho} / d t = \hat{\hat{\mathcal{L}}}\,\hat{\rho}$.

	\vspace{1cm}
	\section{Transport theory in superconducting dissipative media: quantum-field theory approach}\label{sec3}
	
	\subsection{Generalization of Meir-Wingreen formula for the device with dissipation}\label{sec3.1}
	
	In order to characterize charge and energy transport for the system described by the GSKL Eq. \eqref{Lindblad_eq} we need to derive expressions for the corresponding currents
	in the $i$-th lead, $I_{i,c}=\langle\hat{I}_{i,c}\left(t\right) \rangle =e\langle\dot{\hat{n}}_{i} \rangle$ (where $\hat{n}_{i} =\sum_{k}\hat{d}_{ik}^{+}\hat{d}_{ik}$) and $I_{i,\varepsilon}=\langle\hat{I}_{i,\varepsilon}\left(t\right) \rangle =\langle\dot{\hat{H}}_{i} \rangle$. We benefit from the following relations \cite{visuri-22}
	\begin{eqnarray}\label{I1}
		&&\left\langle\dot{\hat{n}}_{i}\right\rangle = \frac{d}{dt} {\rm Tr}\, \{\hat{n}_{i} \hat{\rho}\left(t\right)\}=-i\left\langle[\hat{n}_{i} ,\hat{H}]\right\rangle,\\
		&&\left\langle\dot{\hat{H}}_{i}\right\rangle = \frac{d}{dt} {\rm Tr}\, \{\hat{H}_{i} \hat{\rho}\left(t\right)\}=-i\left\langle[\hat{H}_{i} ,\hat{H}]\right\rangle.\label{I1H}
	\end{eqnarray}
	They hold due to the absence of the leads' operators in the jump operators $\hat{L}_{v}$ \eqref{Lvn}. In other words, as in the dissipationless regime, the currents are determined by the unitary evolution of the density matrix. It implies that one can use exactly the same counting fields for the charge and energy currents as in the case of unitary dynamics. 
	Taking this remark into account, we employ the quantum-field theory approach to calculate the averages of charge and energy current operators $I_{i,c}$ and $I_{i,\varepsilon}$ using the action corresponding to Eq. \eqref{Lindblad_eq} on the Keldysh contour \cite{sieberer-16}. Then, considering fields in a basis of coherent states \cite{kamenev-23b} and after the Larkin-Ovchinnikov rotation \cite{larkin-75} action can be expressed in the following form: 
	\begin{widetext}
		\begin{eqnarray}
			S_{\delta}=\frac{1}{2}\int\limits_{-\infty}^{+\infty}dt\Biggl\{\overline{\Psi}\left(i\partial_{t}\cdot I_{4N}-i\mathcal{L}_{s}\right)\Psi&+&\sum\limits_{j=1}^{N_{L}}\sum\limits_{k}\Bigl[\overline{\Phi}_{jk}\left(i\partial_{t}\cdot I_{4}-H_{jk}\right)\Phi_{jk} \Bigr.\Biggr.
			\notag 
			\\
			&-& \Bigl.\Biggl.\overline{\Phi}_{jk}T_{jk}\left(S_{z}-ia_{j}\xi_{jk}^{\zeta_{\delta}}S_{x}\right)\Psi-\overline{\Psi}\left(S_{z}+ia_{j}\xi_{jk}^{\zeta_{\delta}}S_{x}\right)T_{jk}^{+}\Phi_{jk}\Bigr]\Biggr\},~\delta=c,\varepsilon.\label{S}
		\end{eqnarray} 
	\end{widetext}
	\noindent
	In Eq. \eqref{S} we introduced $4N$-component Keldysh-Nambu spinors to describe quasiparticles in the superconducting part of the system
	\begin{gather}
		\Psi=\left(\Psi^{1}, \overline{\Psi}^{2}, \Psi^{2}, \overline{\Psi}^{1}\right)^{T}, \quad
		\overline{\Psi}=\left({\overline{\Psi}^{1}}^{T}, {\Psi^{2}}^{T}, {\overline{\Psi}^{2}}^{T}, {\Psi^{1}}^{T}\right), \notag\\
		\Psi^{i}=\left(\psi_{1}^{i},\dots,\psi_{N}^{i}\right)^{T},\quad \overline{\Psi}^{i}=\left(\overline{\psi}_{1}^{i},\dots,\overline{\psi}_{N}^{i}\right)^{T}, \label{Psi}
	\end{gather} 
	where $i=1,2$, 
	and $4$-component spinors for the $j$-th lead,
	\begin{equation}\label{Psik}
		\Phi_{jk}{=}\left(\phi_{jk}^{1},\overline{\phi}_{jk}^{2},\phi_{jk}^{2},\overline{\phi}_{jk}^{1}\right)^{T},\quad
		\overline{\Phi}_{jk}{=}\left(\overline{\phi}_{jk}^{1},\phi_{jk}^{2},\overline{\phi}_{jk}^{2},\phi_{jk}^{1}\right) .
	\end{equation}
	In turn, the spinors 
	involve Grassmann variables
	$\psi_{n}^{i},~\phi_{jk}^{i}$ and $\overline{\psi}_{n}^{i},~\overline{\phi}_{jk}^{i}$, $i=1,2$. The parameter $\zeta_{\delta}$ in the right hand side of Eq. \eqref{S} 
	is determined by 
	the index $\delta$: 
	\begin{eqnarray}\label{zeta}
		\zeta_{\delta} = \left\{\begin{array}{*{20}{c}} 
			1,~\delta=\varepsilon; \\
			0,~\delta=c ,  \end{array} \right.
	\end{eqnarray}
	Also it is convenient to define $\bar{\zeta}_{\delta}=1-\zeta_{\delta}$.
	
	We note that to obtain the formula of currents in the lead we have extracted  from the Liouvillian in Eq. \eqref{S} the terms containing degrees of freedom of the 
	reservoirs. The remaining part, 
	$\mathcal{L}_{s}$, describes the superconducting dissipative system alone. $\mathcal{L}_{s}$ has a standard block structure in the Keldysh space and 
	has been found earlier in Ref. \cite{thompson-23},
	\begin{eqnarray}
		&&\mathcal{L}_{s}^{R,A} = -ih_{s}\mp\sum\limits_{v}Q_{v}/2,\label{LRA}\\
		&&\mathcal{L}_{s}^{K} = -\sum\limits_{v}D_{v}.\label{LK}
	\end{eqnarray}
	The contribution from 
	the $v$-th bath to the retarded and advanced components of $\mathcal{L}_{s}$ are given by a matrix $Q_{v}$, see Eq. \eqref{Qf} in Appendix \ref{appA}. In turn, the Keldysh component is determined by a matrix $D_{v}$, see Eq. \eqref{Qf}. We emphasize that  $Q_{v}$ and $D_{v}$ are the Bogoliubov-de-Gennes matrices corresponding to the operators $\left\{\hat{L}_{v}^{+},\hat{L}_{v}\right\}$ and $\left[\hat{L}_{v}^{+},\hat{L}_{v}\right]$ in agreement with the operator definition of retarded, advanced and Keldysh Green's functions, respectively \cite{arseev-15}. 
	
	The degrees of freedom of the leads are contained in the last three terms of Eq. \eqref{S}. The first of these terms describes isolated 
	reservoirs. The other two terms characterize tunneling between the 
	leads and the superconductor 
	and are proportional to the tunneling coefficient matrices $T_{jk}$ ($T_{jk}^{+}$), which are direct sums of the rows (columns) of the parameters $t_{jkn}$, $n=1,\dots,N$,
	\begin{eqnarray}
		&&T_{jk}=\underline{t}_{jk}^{T}\oplus \underline{t}_{jk}^{+}\oplus \underline{t}_{jk}^{T}\oplus \underline{t}_{jk}^{+}\in \mathbb{C}^{4\times4N}, \notag \\
		&&\underline{t}_{jk}=\left(t_{jk1},\dots,t_{jkN}\right)^{T}.\label{Tjk}
	\end{eqnarray}
	The matrices $S_{x,z}$ are constructed using the standard Pauli matrices $\sigma_{x}$ and $\sigma_{z}$: $S_{x}=\sigma_{x}\otimes I_{2N}$; $S_{z}=I_{2} \otimes \tau_{z}$; $\tau_{z}=\sigma_{z} \otimes I_{N}$. The terms proportional to quantum components of sources, $a_{j}$, are inserted to facilitate calculation of the physical observables. Thus, the total action can be split as $S_{\delta}=S_{0}+S_{\delta,a}$.
	
	It is worthwhile to mention that the averaging in the Green's function definition, $G_{s}=-i\left\langle\Psi\overline{\Psi}\right\rangle$, is performed with respect to the bare action $S_{0}$ \eqref{Seff}. Then, the corresponding density matrix is Gaussian unless the non-equilibrium steady state (NESS) is degenerate \cite{prosen-08,prosen-10}. However, following the results of Ref. \cite{shustin-25} one can 
	check that an arbitrary number of zero-energy excitations in the open system renormalizes the density matrix in such a way that it does not 
	affect the averages of single-particle observables (see Sec.\ref{sec5.2}).
	
	After a series of standard transformations detailed in Appendix \ref{appA}, we obtain the following expression for the stationary charge and energy currents in the $j$-th lead:
	\begin{equation}\label{I_MW}
		I_{j,\delta} {=} \frac{i e^{\bar{\zeta}_{\delta}}}{4}\int\limits_{-\infty}^{+\infty}\frac{d\omega}{2\pi} \omega^{\zeta_{\delta}}\Tr\left\{\tau_{z}^{\bar{\zeta}_{\delta}}\Gamma_{j}\left[G_{s}^{K}{-}F_{j}\left(G_{s}^{R}{-}G_{s}^{A}\right)\right]\right\}.
	\end{equation}
	
	For $\zeta_{\delta=c}=0$ and $N_{L}=2$ Eq. \eqref{I_MW} provides a generalization of 
	the Meir-Wingreen formula \cite{meir-92} for the charge current through the superconducting device with dissipation. 
	Eq. \eqref{I_MW} describes the charge current in the $j$-th lead, characterized by the matrices of the reservoirs' distribution functions, $F_{j}$, and broadening parameters, $\Gamma_{j}$, see Eqs. \eqref{Fi}-\eqref{Gammaih}. We emphasize that although in this paper we consider the case of non-interacting quasiparticles in the superconductor, Eq. \eqref{I_MW} is applicable for interacting system as well (similar to the original Meir-Wingreen formula) \cite{meir-93,jauho-94}.  
	Moreover, we suppose that the many-body effects can be caused not only by the well-known internal mechanisms (e.g. electron-electron and electron-phonon interactions) but also by the bath. For example, quadratic jump operators,
		$\hat{L}_{nm} \sim \hat{c}_{n}\hat{c}_{m}$, might mimic many-body effects in the Keldysh effective action.

\begin{figure*}[!htb]
		\begin{center}
			\includegraphics[width=0.5\textwidth]{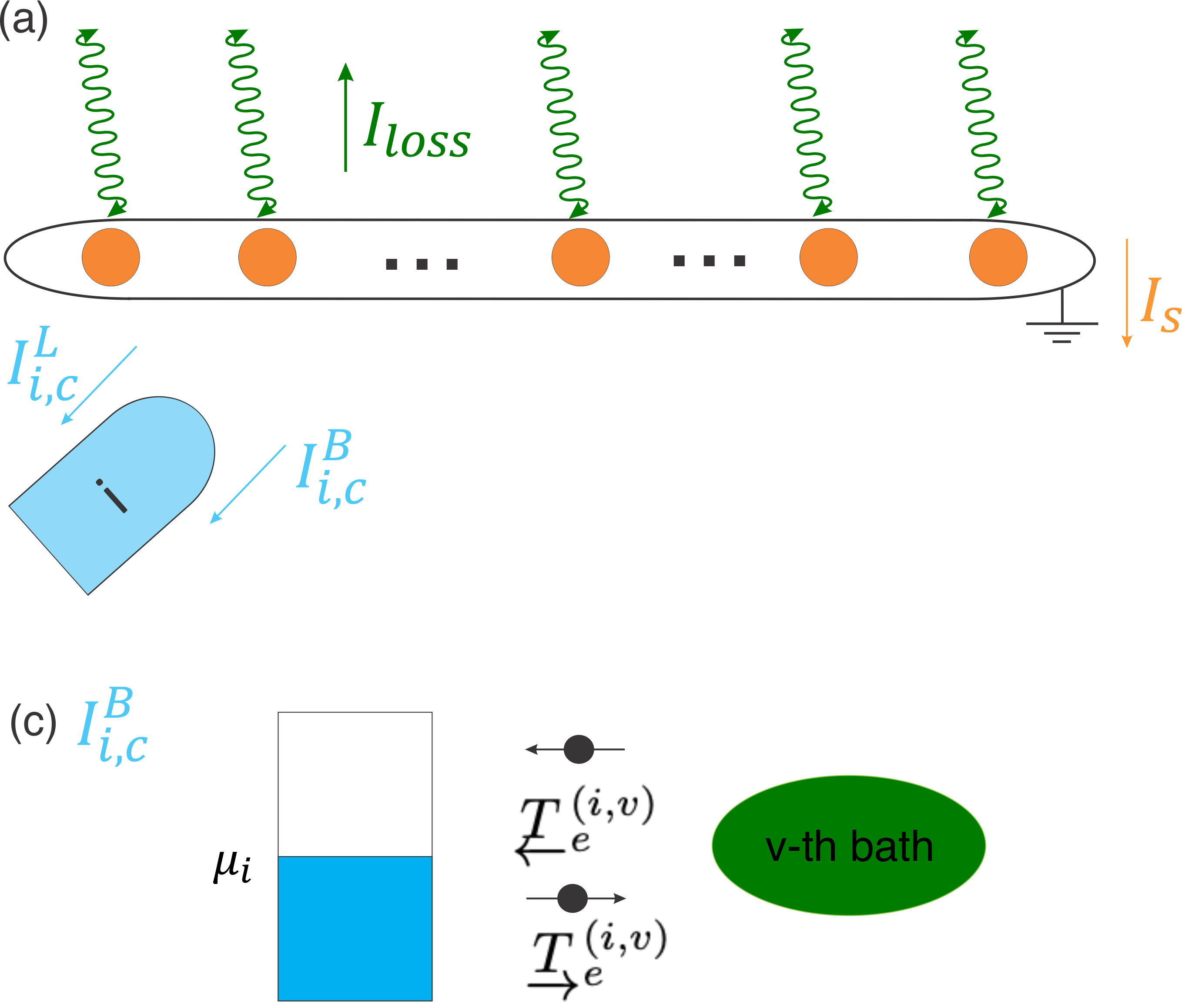}
			\includegraphics[width=0.45\textwidth]{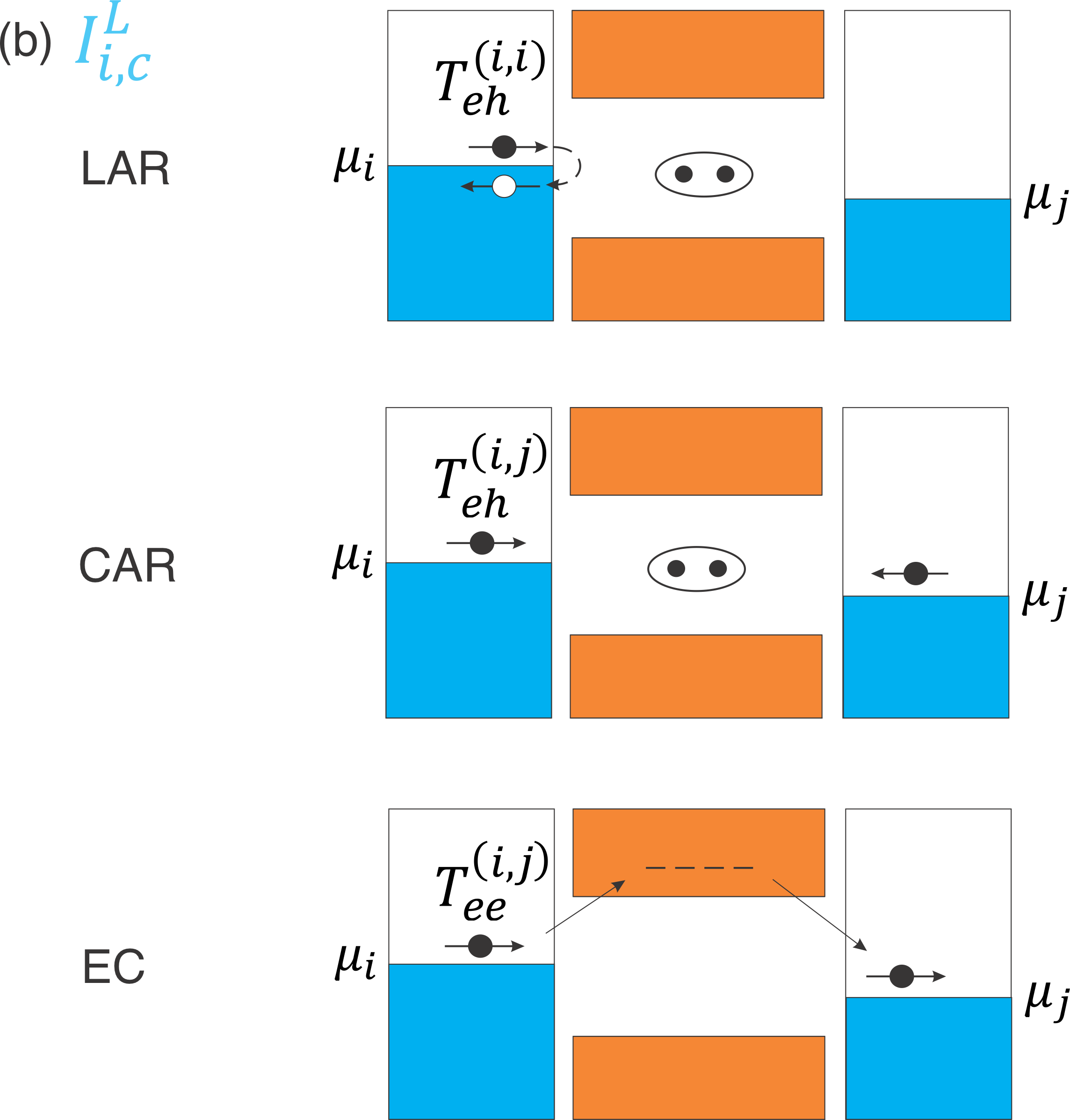}
			\caption{\label{fig2abc} Transport processes in a dissipative superconducting system. (a) The charge current in the $i$-th lead is determined by the contributions from all leads, $I_{i,c}^{L}$, and all baths, $I_{i,c}^{B}$. To satisfy the Kirchhoff's rule in the system one has to take into account both the supercurrent in the directly grounded superconducting device, $I_{s}$, and the loss current, $I_{loss}$, in the total Markovian-type reservoir. (b) There are three transport processes contributing to the current $I_{i,c}^{L}$. Local Andreev reflection (LAR) of an electron incident on the superconductor from the $i$-th lead results in the appearance of a hole in the same lead but moving in the opposite direction, while Cooper pair transmits into the superconductor. Probability of such a process is $T_{eh}^{(i,i)}$. In the case of crossed Andreev reflection (CAR) the electrons comprising the transmitted Cooper pair are taken from different leads ($i \neq j$) and the corresponding probability is $T_{eh}^{(i,j)}$. An electron can also co-tunnels (EC) from $i$-th to $j$-th lead via a virtual state in the superconductor with the probability $T_{ee}^{(i,j)}$. (c) The contribution of $v$-th bath to the current $I_{i,c}^{B}$ represents a superposition of electron inflow and outflow processes between this subsystem and the $i$-th lead with the rates $\underrightarrow{T}_{e}^{(i,v)}$ and $\underleftarrow{T}_{e}^{(i,v)}$, respectively.} 
		\end{center}
	\end{figure*}

	Information about both the coupling to external reservoirs and the internal processes within the device is encoded in the Green's functions $G_{s}^{R,A}$ (retarded and advanced) and $G_{s}^{K}$ (Keldysh), which satisfy the Dyson and Keldysh equations, respectively. As will be shown in Section \ref{sec3.3}, even when accounting for the supercurrent $I_{s}$, Kirchhoff's rule is violated due to the presence of baths, i.e., $\sum\limits_{j}I_{j,c}-I_{s}\neq0$ (see Fig.\ref{fig2abc}a).
	
	At first glance, Eq. \eqref{I_MW} seems to be similar to the main result of Ref. \cite{meir-92}. However, dissipation leads to a number of important subtleties, 
	which will be discussed below. 
	The Fourier transform of the inverse Green's function matrix, corresponding to $S_{0}$ takes the following form
	\begin{widetext}
		\begin{equation}\label{iG}
			G_{s}^{-1}\left(\omega\right) = 
			\left( \begin{array}{*{20}{c}}
				\omega I_{2N}-h_{s}-\sum\limits_{j}\Lambda_{j}\left(\omega\right)+\frac{i}{2}[\sum\limits_{j}\Gamma_{j}\left(\omega\right)+\sum\limits_{v}Q_{v}] & i[\sum\limits_{j}\Gamma_{j}\left(\omega\right)F_{j}\left(\omega\right)+\sum\limits_{v}D_{v}] \\
				0 & \omega I_{2N}-h_{s}-\sum\limits_{j}\Lambda_{j}\left(\omega\right)-\frac{i}{2}[\sum\limits_{j}\Gamma_{j}\left(\omega\right)+\sum\limits_{v}Q_{v}] \end{array} \right).
		\end{equation}
	\end{widetext}
	\noindent
	The effect of the reservoirs 
	on the system is expressed through the self-energy matrices. In the diagonal blocks, they contain the Lamb shifts \eqref{Lambda} and the level broadening, see Eqs.~\eqref{Gammaieh}-\eqref{Gammaih}: $\Sigma^{R,A}\left(\omega\right){=}\sum_{j}\left(\Lambda_{j}\left(\omega\right){+}i\Gamma_{j}\left(\omega\right)/2\right)$. The Keldysh component of the Green's function carries information about the distribution functions of the leads, Eq.~\eqref{Fi}: $\Sigma^{K}\left(\omega\right){=}i\sum_{j}\Gamma_{j}\left(\omega\right)F_{j}\left(\omega\right)$. 
	
	It is important to emphasize the fundamental difference between the self-energy blocks of the baths and those of the leads in the matrix \eqref{iG}. The former, unlike the latter, are local in time, as the GKSL equation is derived within the Markovian approximation. This statement remains valid even in the popular wide-band limit for the leads, where $\Gamma_{j}\neq\Gamma_{j}\left(\omega\right)$. In that case, the frequency dependence still persists in the Keldysh block of the self-energy function coming from the leads.
	
	From the form of \eqref{iG}, it follows that the Dyson and Keldysh equations acquire a modified form (compared to those considered in traditional non-equilibrium non-dissipative problems) due to the contribution of the baths, namely:
	\begin{gather}
		G_{s}^{R} {=} \Bigl(\omega I_{2N}{-}h_{s}{-}\sum\limits_{j}\Lambda_{j}{+}\frac{i}{2}\Bigl[\sum\limits_{j}\Gamma_{j}{+}\sum\limits_{v}Q_{v}\Bigr]\Bigr)^{-1}, \label{eqD}\\
		G_{s}^{K} {=} {-}iG_{s}^{R}\Bigl(\sum\limits_{j}\Gamma_{j}F_{j}{+}\sum\limits_{v}D_{v}\Bigr)G_{s}^{A}, \,\,G_{s}^{A} {=} \left(G_{s}^{R}\right)^{+} .\label{eqK}
	\end{gather}
	
	Given the structure of \eqref{eqD} and \eqref{eqK}, the currents $I_{i,\delta}$ can be formally decomposed into contributions from leads and baths, i.e., (see Fig.\ref{fig2abc}a)
	\begin{eqnarray}\label{ILB}
		I_{i,\delta}=I_{i,\delta}^{L}+I_{i,\delta}^{B}. 
	\end{eqnarray}
	Then, taking into account the symmetry properties of the blocks of the retarded Green's function and the broadening function matrices, these components can be written as (see discussions below Eq.\,\eqref{ILB_app} in Appendix \ref{appA})
	\begin{align}
		I_{i,\delta}^{L} &=  e^{\bar{\zeta}_{\delta}}\sum\limits_{j}\sum\limits_{\alpha=e,h}\int\limits_{-\infty}^{+\infty}\frac{d\omega}{2\pi} \omega^{\zeta_{\delta}}T_{e\alpha}^{(i,j)}\left(n_{ie}-n_{j\alpha}\right)\nonumber\\
		&\hspace{8em} = I_{i,\delta}^{LAR}+I_{i,\delta}^{EC}+I_{i,\delta}^{CAR},\label{IfinC}\\
		I_{i,\delta}^{B} &{=} e^{\bar{\zeta}_{\delta}}\sum\limits_{v}\int\limits_{{-}\infty}^{{+}\infty}\frac{d\omega}{2\pi} \omega^{\zeta_{\delta}}\left[\underrightarrow{T}_{e}^{(i,v)}n_{ie}{-}\underleftarrow{T}_{e}^{(i,v)}\left(1{-}n_{ie}\right)\right].\label{IfinB}
	\end{align}
	From Eq. \eqref{IfinC}, it follows that the contribution of electron-like ($\alpha=e$) or hole-like ($\alpha=h$) carriers 
	of the $j$-th 
	reservoir to the current in 
	the $i$-th lead is determined by an effective transmission coefficient (see Fig.\ref{fig2abc}b), 
	\begin{eqnarray}
		&&T_{e\alpha}^{(i,j)}=\Tr\left\{\Gamma_{ie}G_{e\alpha}\Gamma_{j\alpha}G_{e\alpha}^{+}\right\},\label{Teff}
	\end{eqnarray}
	where $G_{e\alpha}$ denotes a block of the retarded Green's function matrix, $G_{s}^{R}=\left( \begin{array}{*{20}{c}}
		G_{ee} & G_{eh} \\
		G_{he} & G_{hh} \\
	\end{array} \right)$ in the Nambu space. The electron (hole) distribution function in the $j$-th reservoir is given by $n_{je(h)}=\left(1+e^{\frac{\omega\mp eV_{j}}{T+T_{j}}}\right)^{-1}$, where $T$ - equilibrium temperature, $T_{j}$ - deviation from $T$ in the $j$-th lead. 

	For $i = j$, the contribution to $I_{i,\delta}^{L}$ comes only from the local Andreev reflection processes \cite{andreev-64}, $I_{i,\delta}^{LAR}$, (see Fig.\ref{fig2abc}b)
	\begin{equation}
		I_{i,\delta}^{LAR} =  e^{\bar{\zeta}_{\delta}}\int\limits_{-\infty}^{+\infty}\frac{d\omega}{2\pi} \omega^{\zeta_{\delta}}T_{eh}^{(i,i)}\left(n_{ie}-n_{ih}\right).\label{ILAR}
	\end{equation}
	For $i \neq j$, in the general case ($|V_{i}| \neq |V_{j}|$), there are contributions to the current from both the elastic electron cotunneling ($\alpha = e$), $I_{i,\delta}^{EC}$, \cite{falci-01} and the crossed Andreev reflection ($\alpha = h$), $I_{i,\delta}^{CAR}$, \cite{byers-95,deutscher-00}
	\begin{equation}\label{IETCAR}
		\begin{split}
			&I_{i,\delta}^{EC} {=}  e^{\bar{\zeta}_{\delta}} \sum\limits_{j\neq i}\int\limits_{-\infty}^{+\infty}\frac{d\omega}{2\pi} \omega^{\zeta_{\delta}}T_{ee}^{(i,j)}\left(n_{ie}{-}n_{je}\right){=}\sum\limits_{j\neq i}I_{ij,\delta}^{EC},\\
			&I_{i,\delta}^{CAR} {=}  e^{\bar{\zeta}_{\delta}}\sum\limits_{j\neq i} \int\limits_{-\infty}^{+\infty}\frac{d\omega}{2\pi} \omega^{\zeta_{\delta}}T_{eh}^{(i,j)}\left(n_{ie}{-}n_{jh}\right){=}\sum\limits_{j\neq i}I_{ij,\delta}^{CAR}.
		\end{split}
	\end{equation}
	It should be emphasized that expressions \eqref{IfinC}, \eqref{ILAR}, \eqref{IETCAR}, 
	formally has a structure similar to the known results for dissipationless transport in N/S/N structures \cite{byers-95,falci-01,chtchelkatchev-03,morten-06}. However, a key difference here is the renormalization of $T_{e\alpha}^{(i,j)}$ \eqref{Teff} via the Dyson equation \eqref{eqD}, which now includes corrections due to the dissipative baths.
	
	
	As seen from expression \eqref{IfinB}, the contribution of the $v$-th bath to $I_{i,\delta}^{B}$ represents a superposition of electron inflow and outflow processes between this subsystem and the $i$-th reservoir (see Fig.\ref{fig2abc}c). The corresponding rates are given by
	\begin{equation}\label{Te_rates}
		\begin{split}
			\underrightarrow{T}_{e}^{(i,v)} & =\Tr\left\{\Gamma_{ie}\underrightarrow{b}_{ev}\underrightarrow{b}_{ev}^{+}\right\}, \\
			\underleftarrow{T}_{e}^{(i,v)} & =\Tr\left\{\Gamma_{ie}\underleftarrow{b}_{ev}\underleftarrow{b}_{ev}^{+}\right\},
		\end{split}
	\end{equation}
	where
	\begin{equation}\label{be}
		\underrightarrow{b}_{ev}=G_{ee}\,\underline{\mu}_{v}^{*}+G_{eh}\,\underline{\nu}_{v}^{*},~\underleftarrow{b}_{ev}=G_{ee}\,\underline{\nu}_{v}+G_{eh}\,\underline{\mu}_{v}. 
	\end{equation}
	Thus, it follows  from Eqs. \eqref{Te_rates} and \eqref{be}, that, as in the case of the reservoirs, the rates $\underrightarrow{T}_{e}^{(i,v)}$ and $\underleftarrow{T}_{e}^{(i,v)}$ are determined by both normal and Andreev processes. Taking into account the expressions in Eq. \eqref{be}, one can conclude that for a Hermitian or anti-Hermitian $v$-th bath, $\hat{L}_{v} = \pm \hat{L}_{v}^{+}$, the inflow and outflow rates are equal, $\underrightarrow{T}_{e}^{(i,v)}=\underleftarrow{T}_{e}^{(i,v)}$.
	
	We find that when the voltage sign at the $i$-th lead is reversed, in the general case, only the component associated with local Andreev reflection processes \eqref{ILAR} exhibits the symmetry, i.e., $I_{i,\delta}^{LAR}(-V_{i}) = -I_{i,\delta}^{LAR}(V_{i})$.  However, similar to the dissipationless case if the temperatures in the $i$- and $j$-th leads are the same, $T_j=T_i$,  the corresponding summands in Eq. \eqref{IETCAR} can disappear,  
	\begin{equation}
		\begin{split}
			& V_{j} = V_{i}:~~I_{ij,\delta}^{EC}=0,~~I_{ij,\delta}^{CAR}(-V_{i}) = -I_{ij,\delta}^{CAR}(V_{i}) , \\
			& V_{j} = -V_{i}:~~I_{ij,\delta}^{CAR}=0,~~I_{ij,\delta}^{EC}(-V_{i}) = -I_{ij,\delta}^{EC}(V_{i}).
		\end{split}
		\label{IETCAR_sym}
	\end{equation}
	Simultaneously, it is easy to see from Eq. \eqref{IfinB} that in the cases under consideration, $I_{i,\delta}^{B}(-V_{i}) \neq -I_{i,\delta}^{B}(V_{i})$ unless the coupling with all baths is weak (see Sec. \ref{sec5}).
	
	\subsection{Thermoelectric effects in the linear response}\label{sec3.2}
	
	Restricting to first-order terms in the expansion of the Fermi functions, we obtain 
	\begin{equation}
		n_{ie}-n_{j\alpha}\approx\left(-\frac{\partial n_{0}}{\partial\omega}\right) \cdot\left(e\Delta V_{e\alpha}^{(i,j)}+\frac{\omega}{T}\Delta T^{(i,j)}\right) ,    
	\end{equation}
	where $V_{je(h)}=\mp V_{j}$, $n_{0}=\left(1+e^{\omega/T}\right)^{-1}$ stands for the Fermi-Dirac distribution function, $\Delta V_{e\alpha}^{(i,j)}=V_{j\alpha}-V_{ie}$ and $\Delta T^{(i,j)}=T_{i}-T_{j}$ are differences of voltages and temperatures in the $i$-th and $j$-th leads, respectively. Then, the relation between the $L$-contribution of the current 
	in the $i$-th lead, $\left(I_{i,c}^{L},I_{i,\varepsilon}^{L}\right)^{T}$, and the vector of thermodynamic forces acting between the $i$-th and $j$-th leads, $\left(\Delta V_{e\alpha}^{(i,j)},\Delta T^{(i,j)}\right)^{T}$, is determined by the Onsager matrices $L_{e\alpha}^{(i,j)}$,
	\begin{eqnarray}\label{Lij}
		&&\left( \begin{array}{*{20}{c}}
			I_{i,c}^{L} \\
			I_{i,\varepsilon}^{L} \end{array} \right)=\sum\limits_{j}\sum\limits_{\alpha=e,h}\left( \begin{array}{*{20}{c}}
			L_{c,V}^{(i,j)} & L_{c,T}^{(i,j)} \\
			L_{\varepsilon,V}^{(i,j)} & L_{\varepsilon,T}^{(i,j)}
		\end{array} \right)_{\alpha}\left( \begin{array}{*{20}{c}}
			\Delta V_{e\alpha}^{(i,j)} \\
			\Delta T^{(i,j)} \end{array} \right),~~~~~~\\
		&&L_{\delta,V,\alpha}^{(i,j)}=e^{\bar{\zeta}_{\delta}+1}\int\limits_{-\infty}^{+\infty}\frac{d\omega}{2\pi} \omega^{\zeta_{\delta}}\left(-\frac{\partial n_{0}}{\partial\omega}\right)T_{e\alpha}^{(i,j)},\label{CV}\\
		&&L_{\delta,T,\alpha}^{(i,j)}=e^{\bar{\zeta}_{\delta}}\int\limits_{-\infty}^{+\infty}\frac{d\omega}{2\pi} \frac{\omega^{\zeta_{\delta}+1}}{T}\left(-\frac{\partial n_{0}}{\partial\omega}\right)T_{e\alpha}^{(i,j)}.\label{CT}
	\end{eqnarray} 
	As it follows from Eqs. \eqref{Lij}-\eqref{CT},  
	the Onsager relation holds, $L_{\varepsilon,V,\alpha}^{(i,j)} = T  L_{c,T,\alpha}^{(i,j)}$, for given pairs $(i,e)$ and $(j,\alpha)$. 
	
	Using Eq. \eqref{IfinB}, we derive (in the linear approximation) the relation between the $B$-contribution of the current
	in the $i$-th lead, $\left(I_{i,c}^{B},~I_{i,\varepsilon}^{B}\right)^{T}$, and the vector of thermodynamic forces, $\left(-V_{ie},~T_{i}\right)^{T}$:
	\begin{equation}\label{LBij}
		\left( \begin{array}{*{20}{c}}
			I_{i,c}^{B} \\
			I_{i,\varepsilon}^{B} \end{array} \right){=}\sum\limits_{v}\left[\left( \begin{array}{*{20}{c}}
			B_{c,V}^{(i,v)} & B_{c,T}^{(i,v)} \\
			B_{\varepsilon,V}^{(i,v)} & B_{\varepsilon,T}^{(i,v)}
		\end{array} \right)\left( \begin{array}{*{20}{c}}
			{-}V_{ie} \\
			T_{i} \end{array} \right)\!{+}\!\left( \begin{array}{*{20}{c}}
			I_{0,c}^{(i,v)} \\
			I_{0,\varepsilon}^{(i,v)} \end{array} \right)\right], 
	\end{equation}
	where the Onsager matrices are given as
	\begin{align}
		B_{\delta,V}^{(i,v)}&{=}e^{\bar{\zeta}_{\delta}+1}\int\limits_{-\infty}^{+\infty}\frac{d\omega}{2\pi} \omega^{\zeta_{\delta}}\left({-}\frac{\partial n_{0}}{\partial\omega}\right)\left(\underrightarrow{T}_{e}^{(i,v)}{+}\underleftarrow{T}_{e}^{(i,v)}\right), \notag \\
		B_{\delta,T}^{(i,v)}&{=}e^{\bar{\zeta}_{\delta}}\int\limits_{-\infty}^{+\infty}\frac{d\omega}{2\pi} \frac{\omega^{\zeta_{\delta}+1}}{T}\left({-}\frac{\partial n_{0}}{\partial\omega}\right)\left(\underrightarrow{T}_{e}^{(i,v)}{+}\underleftarrow{T}_{e}^{(i,v)}\right),\notag \\
		I_{0,\delta}^{(i,v)}&{=}e^{\bar{\zeta}_{\delta}}\int\limits_{-\infty}^{+\infty}\frac{d\omega}{2\pi} \omega^{\zeta_{\delta}}\left[\underrightarrow{T}_{e}^{(i,v)}n_{0}{-}\underleftarrow{T}_{e}^{(i,v)}\left(1{-}n_{0}\right)\right]. \label{I0B}
	\end{align} 
	
	Assuming that the electrochemical potentials of the leads 
	are controlled independently and the system is in the thermal equilibrium (i.e. $T_{i} = 0$ for all reservoirs),
	the contributions to the differential conductance of the $i$-th lead, $\partial I_{i} / \partial V_{i} = G_{i} = G^{L}_{i} + G^{B}_{i}$, from the leads and baths in the linear response regime acquire the following form:
	\begin{equation}\label{GL_lr}
		G^{L}_{i} = G_{0}\sum\limits_{j}\sum\limits_{\alpha=e,h}\int\limits_{-\infty}^{+\infty}d\omega\left(-\frac{\partial n_{0}}{\partial\omega}\right)  \left(1+\delta_{ij}\right)T_{e\alpha}^{(i,j)},
	\end{equation}
	and
	\begin{equation}\label{GB_lr}
		G^{B}_{i} = G_{0}\sum\limits_{v}\int\limits_{-\infty}^{+\infty}d\omega \left(-\frac{\partial n_{0}}{\partial\omega}\right)\left[\underrightarrow{T}_{e}^{(i,v)}+ \underleftarrow{T}_{e}^{(i,v)} \right],
	\end{equation}
	where the conductance quantum $G_{0} = e^{2}/h$ arises if the Planck constant $\hbar$ is restored in the denominators of Eqs. \eqref{GL_lr} and \eqref{GB_lr}. We note that the current components determined by the equilibrium distribution function, $I_{0,c}^{(i,v)}$, see Eq. \eqref{I0B}, do not affect the conductance. At low temperatures, $T\to 0$, we find $-\partial n_{0}/\partial \omega \to \delta(\omega)$, and the conductance, as is well known, is determined by quasiparticle transport at the energy $\omega = 0$.
	
	\subsection{Loss current and quantum kinetic equations}\label{sec3.3}
	
	As we mentioned above, neglecting the loss current $I_{loss}$, induced due to the presence of dissipation, gives rise to a violation of the Kirchhoff's rule in the system under consideration, $\sum\limits_{j} I_{j,c} - I_{s} \neq 0$ (see Fig.\ref{fig2abc}a). To determine $I_{loss}$, we write the equation of motion for the average charge in the superconducting part of the system, $\left\langle\hat{Q}\right\rangle=e\sum_{n}\left\langle\hat{N}_{n}\right\rangle$, where $\hat{N}_{n}=\hat{c}_{n}^{+}\hat{c}_{n}$, using the GKSL equation \eqref{Lindblad_eq},
	\begin{align}
		\left\langle\dot{\hat{Q}} \right\rangle =& -i\left\langle[\hat{Q} ,\hat{H}]\right\rangle \notag \\
		&+\frac{1}{2}\sum\limits_{v}\left(\left\langle\hat{L}_{v}^{+}\left[\hat{Q},\hat{L}_{v}\right] \right\rangle-\left\langle\left[\hat{Q},\hat{L}_{v}^{+}\right]\hat{L}_{v} \right\rangle\right).\label{dQdt0}
	\end{align}
	Taking into account the commutation relation $\left[\hat{N}_{n},\hat{L}_{v}\right] = -\mu_{vn}\hat{c}_{n}+\nu_{vn}\hat{c}_{n}^{+}$, the straightforward manipulations with the right hand side of Eq. \eqref{dQdt0} yield
	\begin{align}\notag
		\left\langle\dot{\hat{Q}} \right\rangle & = -\sum\limits_{j}I_{j,c}+2e\Tr\left\{\re\{\Delta G_{eh}^{+-}(t,t)\}\right\}\\
		&+ e\sum\limits_{v}\Tr\left\{i\left(\gamma_{v} G_{ee}^{+-}(t,t)-\tilde{\gamma}_{v}^{*}G_{hh}^{+-}(t,t)\right)\right\}.\label{dQdt}
	\end{align}
	Eq. \eqref{dQdt} is expressed in terms of the blocks of the lesser Green's function matrix $G_{s}^{+-}\left(t,t\right)=-i\left\langle T_{C} \hat{\Psi}\left(t\right)\hat{\Psi}^{+}\left(t\right) \right\rangle$ in the Nambu space, where $\hat{\Psi}=\left(\hat{c}_{1},...,\hat{c}_{N},\hat{c}_{1}^{+},...,\hat{c}_{N}^{+}\right)^{T}$. From this definition, it follows that $N_{n} = -i\left[G_{ee}^{+-}(t,t)\right]_{n,n}$ represents the electron occupation of site $n$, and
	$G_{hh}^{+-}(t,t) = iI_{N} - \left(G_{ee}^{+-}(t,t)\right)^{T}$, i.e., $1 - N_{n} = -i\left[G_{hh}^{+-}(t,t)\right]_{n,n}$ is the probability that site $n$ is unoccupied by electron.
	The anomalous averages are determined by the off-diagonal block $G_{eh}^{+-}(t,t)$, with $\left(G_{he}^{+-}(t,t)\right)^{+} = -G_{eh}^{+-}(t,t)$.
	
	The frequency dependence of the blocks $G_{s}^{+-}$ can be determined based on the Keldysh equation $G_{s}^{K} = -G_{s}^{R} \left(G_{s}^{-1}\right)^{K} G_{s}^{A}$ and the relations between different triangular representations \cite{arseev-15},
	\begin{align}
		G_{s}^{+-}&=\frac{1}{2}\left(G_{s}^{K}+G_{s}^{A}-G_{s}^{R}\right) \notag \\
		&= \frac{1}{2}G_{s}^{R}\left[\left(G_{s}^{-1}\right)^{R}-\left(G_{s}^{-1}\right)^{A}-\left(G_{s}^{-1}\right)^{K}\right]G_{s}^{A}.\label{G+-om}
	\end{align}
	Using the explicit form of the blocks of the inverse Green's function matrix \eqref{iG}, we obtain the following kinetic equations:
	\begin{eqnarray}\label{G+-ab}
		G_{\alpha\beta}^{+-} = &&i\left[\sum\limits_{j}\left(G_{\alpha e}\Gamma_{je}G_{\beta e}^{+}n_{je} + G_{\alpha h}\Gamma_{jh}G_{\beta h}^{+}n_{jh}\right)\right.\nonumber\\
		&&\left.~~~~~~~~~~~~~~~~~~+ \sum\limits_{v}\underleftarrow{b}_{\alpha v}\underleftarrow{b}_{\beta v}^{+}\right],~\alpha,\beta=e,h.~~~~~~
	\end{eqnarray}
	Then, taking into account Eq. \eqref{G+-ab} and the definitions for $\underleftarrow{b}_{\alpha v}$, see Eq. \eqref{beh_app}, one can demonstrate that the last two terms in Eq. \eqref{dQdt} are real (it is also related to the Hermitian property of Eq. \eqref{dQdt}), i.e.,
	\begin{eqnarray}\label{tr_re}
		\begin{split}
			&\Tr\left\{i\gamma_{v} G_{ee}^{+-}\right\}=\Tr\left\{i\gamma_{v} G_{ee}^{+-}\right\}^{*},~\\
			&\Tr\left\{i\tilde{\gamma}_{v}^{*} G_{hh}^{+-}\right\}=\Tr\left\{i\tilde{\gamma}_{v}^{*} G_{hh}^{+-}\right\}^{*}.
		\end{split}
	\end{eqnarray}
	
	The second term in Eq. \eqref{dQdt} represents the current carried by the superconducting condensate, $I_{s} \sim 2e\sum\limits_{n,m} \im\left\{\Delta_{nm} \left\langle \hat{c}_{n} \hat{c}_{m} \right\rangle \right\}$ 
	\cite{blonder-82}. In the steady state, $\left\langle \dot{\hat{Q}} \right\rangle = 0$. Thus, taking into account Eq. \eqref{tr_re}, we obtain that the loss current $I_{loss} = \sum\limits_{j} I_{j,c} - I_{s}$ is equal to
	\begin{equation}\label{Iloss2}
		I_{loss}=-e \im \sum\limits_{v}\int\limits_{-\infty}^{+\infty}\frac{d\omega}{2\pi }\Tr \left [ \gamma_{v} G_{ee}^{+-}(\omega)-\tilde{\gamma}_{v}^{*}G_{hh}^{+-}(\omega) \right ].
	\end{equation}
	In the special case of a non-superconducting quantum dot ($\Delta_{nm} = 0$, $N = 1$) with $N_{L} = 2$ and $N_{B} = 1$, we find $I_{loss} = I_{1,c} + I_{2,c} = -e\gamma N_{1}$ that coincides with a result obtained in Ref. \cite{visuri-22b}. By analogy with Ref. \cite{jin-20}, the loss current, Eq. \eqref{Iloss2}, can also be interpreted as the current flowing into all Markovian-type (dissipative) leads.

	\section{Zero-energy excitations in the presence of wide-band leads and dissipation}
	\label{sec4}
	
	In comparison with closed electron systems the open ones are characterized by two gaps, a spectral gap and a purity gap \cite{bardyn-13,altland-21}. The absence of former in superconducting media is an indication of the existence of zero kinetic modes, superpositions of Majorana modes wave functions stable to the effect of dissipation \cite{shustin-25}. The absence of the purity gap implies a mixed nature of NESS. The tunnel quantum-transport  measurements allow to probe directly the spectral gap  and, in general, the device spectrum which is given by an effective Hamiltonian located in the $G_{s}^{R}$ denominator, see Eq. \eqref{eqD}. For example, the detection of the Majorana state in a setup with a single wide-band lead \cite{law-09,flensberg-10} is based on the observation of the zero-bias conductance peak implying that the effective Hamiltonian 
	contains an eigenvalue with a zero real part and nonzero imaginary part (that determines the peak width and, in other words, the lifetime of the damped Majorana mode) as well as one zero eigenvalue corresponding to a zero kinetic mode (see e.g. Eq.\eqref{eigenval12_pt1_re} at $\underline{l}_{1}^{r,i}=0$). Below, we briefly demonstrate how the presence of an arbitrary number of 
	reservoirs, treated in the wide-band approximation, and baths affects the realization of the zero kinetic modes.
	
	In the wide-band approximation, when analyzing the self-energy functions, one can neglect the Lamb shifts of the system levels caused by the coupling to the Fermi reservoirs and treat the corresponding broadening matrices as frequency-independent, i.e.,
	\begin{eqnarray}
		&&\Lambda_{j}=0,~\Gamma_{je}= 2\pi \underline{t}_{j}^{*}\underline{t}_{j}^{T}\rho_{j}, \Gamma_{jh}=\Gamma_{je}^{T},\label{Gm_WBL}\\
		&&\Sigma_{j}^{R,A}=\mp i\pi\rho_{j}\left(\underline{t}_{j}^{*}\underline{t}_{j}^{T}\oplus\underline{t}_{j}\underline{t}_{j}^{+}\right),\label{Sgm_WBL}
	\end{eqnarray} 
	where $\rho_{j}=\sum\limits_{k}\delta\left(\omega-\xi_{jk}\right)=\const$ denotes the density of states of the $j$-th lead which is assumed to be exactly the same for electron and hole states.
	
	As a result, Eq. \eqref{iG} 
	suggests that the spectral properties of the system in the presence of dissipation are determined by the effective Hamiltonian
	\begin{eqnarray}\label{Heff}
		H_{eff}=h_{s}-\frac{i}{2}\left (\sum\limits_{j}\Gamma_{j}+\sum\limits_{v}Q_{v}\right ).
	\end{eqnarray}
	
	It is instructive to discuss a description of the system in terms of the self-conjugate A, B-type lattice Majorana operators , i.e. $\hat{\gamma}_{An}=\hat{c}_{n}+\hat{c}_{n}^{+}$, $\hat{\gamma}_{Bn}=i\left(\hat{c}_{n}^{+}-\hat{c}_{n}\right)$. It can be derived by the transformation of $H_{eff}$ with the help of the rotation matrix
	\begin{equation}\label{Ufm}
		\mathcal{U}=\left( \begin{array}{*{20}{c}}
			I_{N} & ~~~I_{N} \\
			-iI_{N} & iI_{N} \\
		\end{array} \right) ,\end{equation} 
	such that the spectral properties in the Fock-Liouville space are described by a matrix
	\begin{equation}\label{Xmod}
		\tilde{X}=i\left(\mathcal{U}^{-1}\right)^{+}H_{eff}\mathcal{U}^{-1}=-2A+\sum\limits_{v}\ell_{v}\ell_{v}^{T}+\sum\limits_{j}\Upsilon_{j}\Upsilon_{j}^{T} ,
	\end{equation} 
	where $\left(\mathcal{U}^{-1}\right)^{+}=\mathcal{U}/2$.
	To highlight the presence of the leads, here and below, we use the tilde symbol 
	``\textasciitilde'' 
	to indicate a modification of the quantities of interest as compared to the case of the 
	dissipative system without the leads.
	In Eq. \eqref{Xmod} the matrix $A=-i\left(\mathcal{U}^{-1}\right)^{+}h_s\mathcal{U}^{-1}/2$ is a real skew-symmetric matrix, $A^T=-A$,  representing the Hamiltonian in the space of Majorana operators. The matrix $\ell_{v} = \left(\underline{l}_{v}^{r}~~\underline{l}_{v}^{i}\right)$ characterizes the dissipative field of the $v$-th bath in the Majorana basis. It consists of two column components,
	\begin{equation}
		\underline{l}^{r}_{v}=\frac{1}{2}
		\begin{pmatrix}
			\re (\underline{\mu}_{v}+\underline{\nu}_{v}) \\
			-\im (\underline{\mu}_{v}-\underline{\nu}_{v})
		\end{pmatrix},~~ 
		\underline{l}^{i}_{v}=\frac{1}{2}\begin{pmatrix}
			\im (\underline{\mu}_{v}+\underline{\nu}_{v}) \\
			\re (\underline{\mu}_{v}-\underline{\nu}_{v})
		\end{pmatrix}\label{lri_ferm0}
	\end{equation}
			In each component, the first $N$ elements represent the coupling amplitudes of the bath to Majorana fermions of type $A$, $\underline{l}^{r,i}_{vA}$, while the next $N$ elements correspond to the coupling with Majorana fermions of type $B$, $\underline{l}^{r,i}_{vB}$.
			
			Finally, the last term in Eq. \eqref{Xmod} describes the influence of the leads on the spectrum in the Fock-Liouville space. Analogously to the baths, the contribution of the $j$-th 
			reservoir is determined by the following two-component matrix:
			\begin{equation}\label{Ups}
				\Upsilon_{j}=\left(\underline{t}_{j}^{r}~~\underline{t}_{j}^{i}\right)=\frac{1}{2}\sqrt{2\pi\rho_j}\begin{pmatrix}
					\re \underline{t}_{j} & \im \underline{t}_{j} \\
					-\im \underline{t}_{j} & \re \underline{t}_{j}
				\end{pmatrix} .
			\end{equation}
			As it was done for $\underline{l}^{r,i}_{v}$, it is convenient to organize vectors $\underline{t}_{j}^{r,i}$ in accordance with the $A$ and $B$ sublattices, i.e., 
			\begin{equation}
				\underline{t}_{j}^{r,i}=\begin{pmatrix}
					\underline{t}_{jA}^{r,i} \\
					\underline{t}_{jB}^{r,i}
				\end{pmatrix} .     
			\end{equation}
			
			Note that a direct comparison of the matrix $\tilde{X}$ with the similar one in \cite{shustin-25} requires taking into account the factor $2$, which distinguishes the dissipators of the GKSL equations in these works.
			
			As seen from Eqs. \eqref{lri_ferm0} and \eqref{Ups}, 
			$\Upsilon_{j}$ has the same structure as the dissipative field $\ell_v$ in which $\underline{\mu}_v$ is substituted by $\underline{t}_j$ while $\nu_v$ is identically zero. Physically, the difference between the fields of the leads and the dissipative baths in the Majorana operator representation is related to the fact that the Lindblad operators $\hat{L}_{vn}$ effectively account for both electron and hole creation processes (a superconducting-type bath). If $\underline{\mu}_{v} = 0$ or $\underline{\nu}_{v} = 0$, then the fields $\underline{l}_{v}^{r,i}$ and $\underline{t}_{v}^{r,i}$ are qualitatively identical. It follows that, unlike the case $\hat{L}_{v} = \hat{L}_{v}^{+}$ 
			for which $l_v^i\equiv 0$, the Hermiticity of the tunneling Hamiltonian \eqref{HT} does not lead to $\underline{t}_{j}^{r} = 0$ or $\underline{t}_{j}^{i} = 0$. For example, when $\underline{\mu}_{v} = \underline{\nu}_{v}$ and $\operatorname{Im}\underline{\mu}_{v}=\im \underline{\nu}_{v} = 0$, we have $\underline{l}^{r}_{vA} \neq 0$, $\underline{l}^{r}_{vB} = 0$, and $\underline{l}^{i}_{vA,B} = 0$. In contrast, in the case $\operatorname{Im}{\underline{t}_{j}} = 0$ we find $\underline{t}_{jA}^{r} = \underline{t}_{jB}^{i} = \underline{t}_{j}$  while $\underline{t}_{jB}^{r} = \underline{t}_{jA}^{i} = 0$.
			
			In Ref. \cite{shustin-25}, we have shown that the number of zero eigenvalues, $\beta = 0$, of the matrix $\tilde{X}$ which correspond to the zero-energy excitations 
			is equal to (we assume $N_{M}\geqslant N_{B}+N_{L}$)
			\begin{eqnarray}\label{M0}
				N_0 = \dim \ker \tilde{\mathcal{B}}^T = 
				2N_{M}-\rk\tilde{\mathcal{B}} .
			\end{eqnarray}
			Here the 
			$2N_M{\times}(2N_B{+}2N_L)$ matrix $\tilde{\mathcal{B}}$ describes the 
			hybridization of $N_{M}$ pairs of Majorana modes 
			with $N_{B}$ pairs of bath fields and $N_{L}$ pairs of lead fields,
			\begin{equation}\label{Btilde}
				\tilde{\mathcal{B}} = \begin{pmatrix}
					\mathcal{B}_{B}~~\mathcal{B}_{L}
				\end{pmatrix} = 
				\begin{pmatrix}
					\underline{\chi}_{1} \dots \underline{\chi}_{2N_{M}}    
				\end{pmatrix}^T    
				\begin{pmatrix}
					\ell~~\Upsilon 
				\end{pmatrix}
			\end{equation}
			where we introduce the $2N{\times}2N_B$ matrix $\ell=(\underline{l}_1^r,\underline{l}_1^i,\dots\,,\underline{l}^{r}_{N_B},\underline{l}^{i}_{N_B})$ and the $2N{\times}2N_L$ matrix
			$\Upsilon=(\underline{t}_{1}^{r},\underline{t}_{1}^{i},\dots,\underline{t}_{N_{L}}^{r},\underline{t}_{N_{L}}^{i})$. The vectors 
			$\underline{\chi}_{a}$, $a = 1, \dots, 2N_{M}$, represent the wave functions of Majorana zero energy states in the isolated system, i.e. they solve the following equation $A\underline{\chi}_{a} = 0$.

			When considering the lead fields, it is natural to assume their local nature. In other words, in the common method of tunneling spectroscopy, where different leads are connected, for example, with the opposite ends of one-dimensional structure, we have $t_{1,n} \sim \delta_{1,n}$ and $t_{2,n} \sim \delta_{N,n}$ (a similar structure of $\underline{t}_{j}$ also appears in the scanning tunneling microscopy experiments). Taking this fact into account, the number of the zero-energy states in the open system, $N_0$, is determined by the expression (under the assumption $N_{M}\geq N_{B}+N_{L}$)
			\begin{equation}\label{M}
				N_0= - \, {\rm ind\,} \tilde{\mathcal{B}}_B + \dim \ker \tilde{\mathcal{B}}_B -N_L  ,
			\end{equation} 
			where $-\, {\rm ind\,} \tilde{\mathcal{B}}_B=2(N_{M}-N_{B})$ gives the number of robust zero modes and  $\dim \ker \tilde{\mathcal{B}}_B = 2\min(N_{B}, N_{M}) - \operatorname{rk}\mathcal{B}_{B}$ determines the number of weak zero modes appearing due to fine-tuning of the parameters of the dissipative baths.    
			A detailed derivation of expressions \eqref{M0}-\eqref{M} is presented in Ref. \cite{shustin-25}. In Appendix \ref{appB} we discuss several examples of applying general expression \eqref{M} to the case where $N_{B} + N_{L} \leqslant 2$ and $N_{L} \neq 0$. 
			
			We mention that one can refuse wide-band 
			assumption for the leads and deal with the frequency-dependent $H_{eff}$ (including the lead-induced Lamb shift). In this case, the eigenenergies have to be found via the solution of transcendent equation $\det\left[ H_{eff}\left(\omega\right) - \omega I_{2N}\right]=0$. Apparently, the behavior of the solutions is able to demonstrate intriguing features related to a finite cut-off energy of the  spectrum in the leads and to the explicit form of the dispersion law \cite{visuri-22b}.
			
			\section{Applications}\label{sec5}
			
			\subsection{\label{sec5.1} 
				Standard Kitaev chain with a weakly coupled lead and bath }
			
			Let us apply the general theory developed above to one of the currently popular transport problems concerning superconducting wires. That is the standard Kitaev chain interacting with a single wide-band lead and with bath. We will focus  
			on the charge transfer that implies $\delta=c$; $i,j,v=1$ in Eqs. \eqref{ILB}-\eqref{be} (the results for the thermal transport will be reported elsewhere). We are interested at the bias-voltage-dependence of conductance in the zero-temperature limit $G_{1j}\left(V\right)$, where the index $j=L\left(R\right)$ when the lead is coupled to the leftmost (rightmost) site of the chain. The broadening matrix blocks $\Gamma_{1e}$, $\Gamma_{1h}$ \eqref{Gm_WBL} have the only nonzero element, $\left[\Gamma_{1e}\right]_{ij}=\left[\Gamma_{1h}\right]_{ij}=\Gamma_{1}\delta_{i,1(N)}\delta_{j,1(N)}$ if $j=L\left(R\right)$.
			
\begin{figure*}[!htb]
				\begin{center}
					\includegraphics[width=1.05\textwidth]{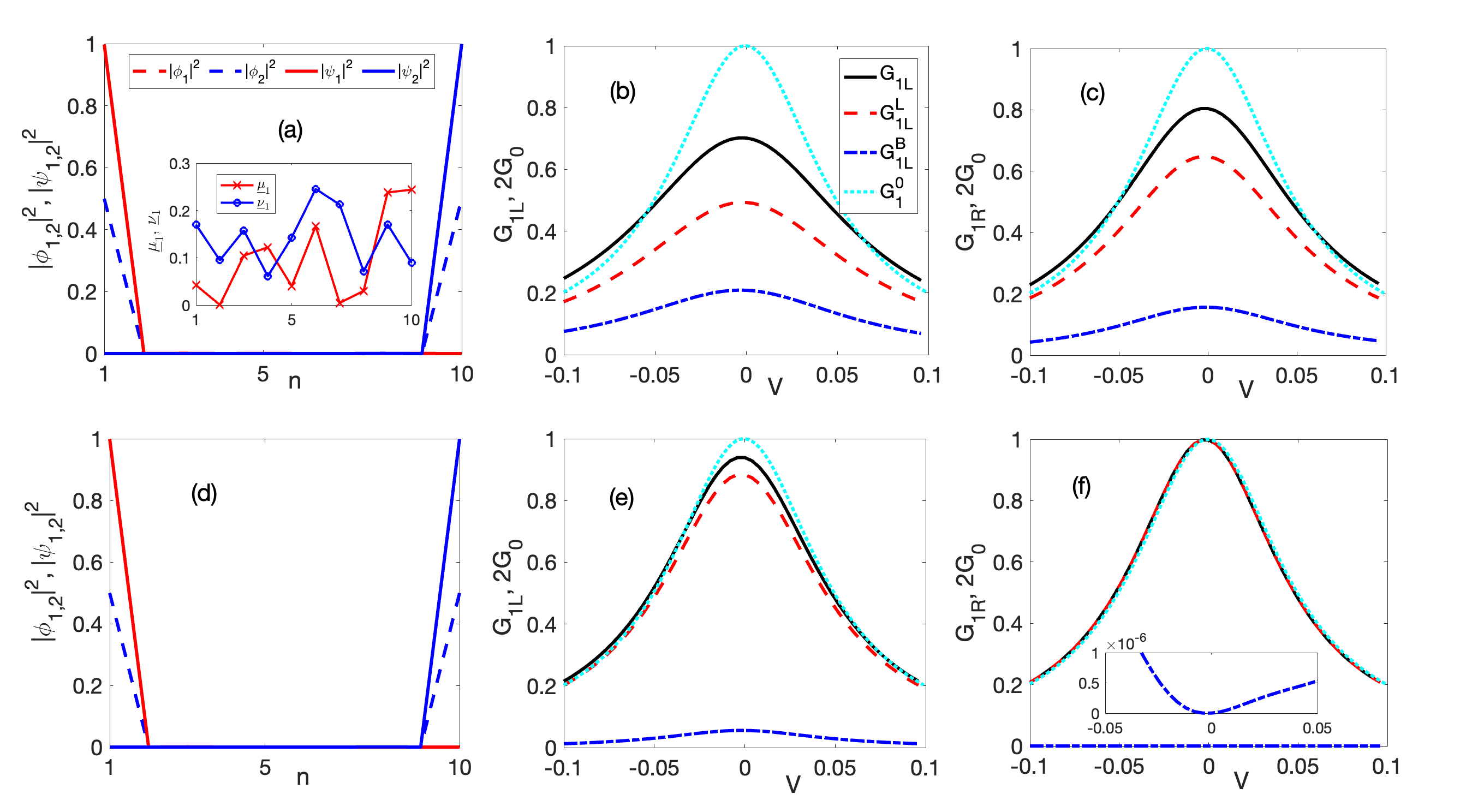}
					\caption{\label{fig3} The effect of a single weak bath with real amplitudes $\underline{\mu}_{1}, \underline{\nu}_{1}$ on the low-energy excitations and local tunnel transport in the standard Kitaev chain model. The top (bottom) row corresponds to the case of a non-Hermitian (Hermitian) bath. (a,d) Spatial distributions of the probability densities of the states in the closed system $|\phi_{1,2}|^2$ (dashed curves) and in the open system $|\psi_{1,2}|^2$, corresponding to energies with zero real part (solid curves). (b(c),e(f)) Voltage dependence of the conductance of lead attached to the left (right) end of the wire, $G_{1L}(V)$ ($G_{1R}(V)$). In panels (b,c,e,f), the dotted curve, $G_{1}^{0}(V)$, corresponds to the situation of the single-lead transport in the absence of dissipation ($\underline{\mu}_{1}=\underline{\nu}_{1}=0$). The red dashed curve shows the contribution from the local Andreev reflection to the conductance $G_{1L,1R}^{L}$, while the blue dash-dotted line represents the contribution from the bath $G_{1L,1R}^{B}$. Inset of panels (a,d): spatial distribution of the dissipative fields $\underline{\mu}_{1},~\underline{\nu}_{1}$. Inset of panel (f): zero-bias dip of $G_{1R}^{B}$ when the bath is Hermitian. We use the following parameters: $N=10$, $\mu=0$, $t=1$, $\Delta=-0.95$, $\Gamma_{1}=0.1$.} 
				\end{center}
			\end{figure*}

			In terms of the Hamiltonian \eqref{Hs}, we set $\xi_{nn}=-\mu$, $\xi_{nm}=t\delta_{m,n-1}=t\delta_{m,n+1}$ and $\Delta_{nm}=-\Delta\delta_{m,n-1}=\Delta\delta_{m,n+1}$ with real parameters $t$ and $\Delta$: $\im t =\im\Delta = 0$.
			Then at the symmetric point, $t=-\Delta$ and $\mu=0$, the skew-symmetric matrix $A$ has the following nonzero elements $A_{n+1,n+N}=-A_{n+N,n+1}=t/2$, $n=1,\dots,N-1$ such that the closed system hosts a single Bogoliubov excitation with zero energy. 
			The wave functions of the corresponding Majorana modes
			\begin{equation}\label{MM12}
				\underline{\chi}_1=\left( \begin{array}{*{20}{c}}
					\underline{\chi}_{1A} \\
					0 \\
				\end{array} \right),~ \underline{\chi}_2=\left( \begin{array}{*{20}{c}}
					0 \\
					\underline{\chi}_{2B} \\
				\end{array} \right)
			\end{equation}
			are localized at the left and right ends of the wire, respectively: $\chi_{1,n}=\delta_{n,1}$ and  $\chi_{2,n}=\delta_{n,2N}$. This scenario is illustrated in Fig.\ref{fig3}a, where the red dashed and and blue dotted lines (the curves coincide) show the spatial distributions of the probability densities for the electron- and hole-like Bogoliubov excitations with zero energy, $|\phi_{1,2}(n)|^2$, where $\underline{\phi}_{1,2} = \frac{1}{2} \left( \begin{array}{*{20}{c}}
				\underline{\chi}_{1A}\pm \underline{\chi}_{2B} \\
				\underline{\chi}_{1A}\mp \underline{\chi}_{2B} \\
			\end{array} \right)$. Deviation from the symmetric point causes spatial oscillations of $|\chi_{1,2}(n)|^2$ in addition 
				with exponential decay into the bulk. As a result, $|\phi_{1,2}(n)|^2$ becomes nonzero at $n\neq1,N$ pressing towards the edges as the system size increases \cite{valkov-22}. One can expect similar behavior of the lowest-energy excitations in the open system, $\psi_{1,2}$, if the bath and lead fields are weak enough (see Eq. \eqref{eigenvec12_pt1_re}). We note in passing that the attempts to detect Majorana modes $\underline{\chi}_{1,2}$ in tunnel transport experiments with hybrid superconducting wires, which have yielded controversial results, have been ongoing for more than a decade.
			
			Taking the derivative with respect to $V$ in the expressions \eqref{IfinC} and \eqref{IfinB}, we find 
			\begin{equation}
				G_{1j} = G_{1j}^{L}+G_{1j}^{B},~j=L,R .\\
			\end{equation}
			Here we split the contributions from the lead and from the bath. They are given as 
			\begin{align}
				G_{1j}^{L} & = G_{0}\left[T_{eh,j}^{\left(1,1\right)}\left(V\right)+T_{eh,j}^{\left(1,1\right)}\left(-V\right)\right]\nonumber\\
				& = G_{0}\Gamma_{1}^{2}\left[|g_{j}\left( V\right)|^2+|g_{j}\left( -V\right)|^2\right],\label{G1V_LAR}\\
				G_{1j}^{B} & = G_{0}\left[\underrightarrow{T}_{e,j}^{(1,1)}\left(V\right)+\underleftarrow{T}_{e,j}^{(1,1)}\left(V\right)\right]\nonumber\\ & = G_{0}\Gamma_{1}\left[\underrightarrow{b}_{j}\left(V\right)+\underleftarrow{b}_{j}\left(V\right) \right], \label{G1V_B}
			\end{align}
			where
			\begin{align}\notag
				& g_{L\left(R\right)}\left(V\right) =\Bigl[G_{eh}\left(V\right)\Bigr]_{1\left(N\right),1\left(N\right)},\\
				& \underrightarrow{b}_{L\left(R\right)} =\Bigl[\underrightarrow{b}_{e1}\left(V\right)\underrightarrow{b}_{e1}^{+}\left(V\right)\Bigr]_{1\left(N\right),1\left(N\right)}, \\
				& \underleftarrow{b}_{L\left(R\right)}=\Bigl[\underleftarrow{b}_{e1}\left(V\right)\underleftarrow{b}_{e1}^{+}\left(V\right)\Bigr]_{1\left(N\right),1\left(N\right)}.
			\end{align}
			Then, the conductance contributions \eqref{G1V_LAR} and \eqref{G1V_B} are determined by blocks of the retarded Green's function, $G_{ee}$ and $G_{eh}$. To find them it is convenient to use the spectral decomposition, 
			\begin{align}\notag
				G_{s}^{R}\left(\omega\right)  =\Bigl[\omega I_{2N} & -H_{eff}\Bigr]^{-1}\nonumber\\
				& = \mathcal{U}^{-1}U \bigoplus_{a=1}^{2N}\left(\omega+2i\beta_{a}\right)^{-1} U^{-1}\mathcal{U},\label{Gr_specdec}
			\end{align}
			where $\beta_{a}=\beta_{a}'+i\beta_{a}''$ stands for an eigenvalue of the non-Hermitian Hamiltonian $\tilde{X}$ \eqref{Xmod}.
			
			The matrix $\mathcal{U}$ is responsible for the transform from the fermion operator basis to the Majorana operator basis in real space and is given by the expression \eqref{Ufm}. The matrices $U$ and $U^{-1}$ consist of the right and left eigenvectors of $\tilde{X}$,
			\begin{align}\notag
				& U=\Bigl[|R_{1}\rangle,\dots,|R_{2N}\rangle\Bigr],~U^{-1}=\Bigl[\langle L_{1}|,\dots,\langle L_{2N}|\Bigr]^T,\\
				& \tilde{X}|R_{a}\rangle=\beta_{a}|R_{a}\rangle,~\langle L_{a}|\tilde{X}=\beta_{a} \langle L_{a}|,~\langle L_{a}|R_{b}\rangle=\delta_{ab}.\label{eigenlr}
			\end{align}
			Their structure in the Majorana representation is 
			\begin{equation}\label{RLaAB}
				|R_{a}\rangle=\begin{pmatrix}
					|r_{aA}\rangle \\
					|r_{aB}\rangle
				\end{pmatrix},~|L_{a}\rangle=\begin{pmatrix}
					|l_{aA}\rangle \\
					|l_{aB}\rangle
				\end{pmatrix}.
			\end{equation}
			
			Next, we assume that the coupling to the lead and bath is weak, i.e. $\Gamma_{1},~\left[Q_{1}\right]_{i,j}\ll |t|,~|\Delta|$. To achieve this regime in the quantum-transport numerical calculations in Fig. \ref{fig3} we set $\Gamma_{1}=0.1$. Also we draw $\mu_{vn}$ and $\nu_{vn}$ from a standard uniform distribution on the open interval $\left(0, 0.25\right)$. A typical spatial dependence of $\mu_{1,n},~\nu_{1,n}\in \mathbb{R}$ is displayed in the inset of Fig. \ref{fig3}a. Thus, in this case the spectrum corrections are weak and still retain a considerable gap $|\beta_{a\neq1,2}|-|\beta_{1,2}|\sim2t$. We are interested at the subgap response, $|eV|\ll2t$, then only first two terms can be left in the decomposition \eqref{Gr_specdec}. Consequently, the following approximation for the blocks of $G_{s}^{R}$ is valid:
			\begin{equation}\label{GeeGeh_lr}
				G_{ee}\approx\frac{1}{2}\sum\limits_{a=1,2}\frac{|r_{a}\rangle \overline{\langle l_{a}|}}{\omega+2i\beta_{a}},~G_{eh}\approx\frac{1}{2}\sum\limits_{a=1,2}\frac{|r_{a}\rangle\langle l_{a}|}{\omega+2i\beta_{a}},
			\end{equation}
			where 
			\begin{equation}
				\label{rala}
				\begin{split}
					|r_{a}\rangle & =|r_{aA}\rangle+i|r_{aB}\rangle,~\langle l_{a}|=\langle l_{aA}|+i\langle l_{aB}| , \\
					\overline{|r_{a}\rangle} & =|r_{aA}\rangle-i|r_{aB}\rangle,~\overline{\langle l_{a}|}=\langle l_{aA}|-i\langle l_{aB}| .
				\end{split}
			\end{equation}
			
			Finally, the contributions $G_{1j}^{L}$ and $G_{1j}^{B}$ become 
			\begin{widetext}
				\begin{align}\notag
					& G_{1L(R)}^{L}=2G_{0}\left(\Gamma_{1}/2\right)^{2}\sum\limits_{a,b=1}^{2}\frac{\left(eV\right)^2+4\beta_{a}\beta_{b}^{*}}{\left(\left(eV\right)^2+4\beta_{a}^2\right)\left(\left(eV\right)^2+4\left(\beta_{b}^{*}\right)^2\right)}\Bigl[|r_{a}\rangle\langle l_{a}|\Bigr]_{1\left(N\right),1\left(N\right)}\cdot\Bigl[|r_{b}\rangle \langle l_{b}|\Bigr]_{1\left(N\right),1\left(N\right)},\\
					&G_{1L(R)}^{B}=2G_{0}\left(\Gamma_{1}/4\right)\sum\limits_{a,b=1}^{2}\frac{\langle L_{a}|\mathcal{U}Q_{1}\mathcal{U}^{-1}|L_{b}\rangle}{(eV+2i\beta_{a})(eV-2i\beta_{b}^{*})}\Bigl[|r_{a}\rangle\langle r_{b}|\Bigr]_{1\left(N\right),1\left(N\right)},\label{GLARGB_fin}
				\end{align}
			\end{widetext}
			\noindent
			where 
			\begin{equation}\label{UQU-1}
				\mathcal{U}Q_{1}\mathcal{U}^{-1}=\left( \begin{array}{*{20}{c}}
					\re Q_{+} & ~-\im Q_{+} \\
					\im Q_{-} & \re Q_{-} \\
				\end{array} \right)=4\ell_{1}\ell_{1}^{T},
			\end{equation}
			and $Q_{\pm}=\gamma_{1}+\tilde{\gamma}_{1}\pm\eta_{s1}$.
			
			The eigenenergies $\beta_{1,2}$ and eigenvectors $|R_{1,2}\rangle$, $\langle L_{1,2}|$ can be found within the first-order perturbation theory for a perturbation operator $M=\ell_{1}\ell_{1}^{T}+\Upsilon_{1}\Upsilon_{1}^{T}$. As the lowest-energy excitations in the closed system are twofold degenerate the corrections to the energies are eigenvalues of the perturbation operator restricted to the subspace of the degenerate states $\underline{\chi}_{1,2}$. As a result, one deals with the product $\tilde{\mathcal{B}}\tilde{\mathcal{B}}^{T}$ (see Eq. \eqref{Btilde}) for $N_{B,L}=1$. 
			Then, the eigenvalues $\beta_{1,2}$ are roots of a secular equation $\det\left[\tilde{\mathcal{B}}\tilde{\mathcal{B}}^{T} - \beta I_{2}\right]=0$. For a sake of simplicity, we consider the case of real lead and bath fields. Then, due to the relations \eqref{lri_ferm0} and \eqref{Ups}, the matrix $\tilde{\mathcal{B}}\tilde{\mathcal{B}}^{T}$ is diagonal, $\left[\tilde{\mathcal{B}}\tilde{\mathcal{B}}^{T}\right]_{1,2}=0$. Hence, 
			the energies of the initially degenerate Majorana modes are modified by the bath and lead fields as follows:
			\begin{align}
				& \beta_{1}^{(j)}=\left(\underline{\chi}_{1}^{T}\underline{l}_{1}^{r}\right)^2+\frac{\Gamma_1}{4}\chi_{1,1}^{2}\delta_{j,L},\nonumber\\
				&\beta_{2}^{(j)}=\left(\underline{\chi}_{2}^{T}\underline{l}_{1}^{i}\right)^2+\frac{\Gamma_1}{4}\chi_{2,2N}^{2}\delta_{j,R},~j=L,R.\label{eigenval12_pt1_re}
			\end{align}
			Here the subscript $j=L(R)$ indicates the magnitudes of eigenvalues for the case of a lead coupled to left (right) end of the chain. Also in Eq. \eqref{eigenval12_pt1_re} we used the explicit form of the lead fields \eqref{Ups}, i.e. $t_{1,n}^{r(i)}=\frac{\sqrt{\Gamma_{1}}}{2}\delta_{1(N+1),n}$ ($t_{1,n}^{r(i)}=\frac{\sqrt{\Gamma_{1}}}{2}\delta_{N(2N),n}$) if the lead is attached to the left (right) side of the chain.
			Additionally, the bare wave functions of the Majorana modes remain unperturbed, 
			\begin{equation}
				|R_{1,2}\rangle=\underline{\chi}_{1,2},\label{eigenvec12_pt1_re}
			\end{equation}
			where we have neglected the contributions to $|R_{1,2}\rangle$ from the states above the gap as the same approximation has been used for $G_{s}^{R}$ in Eq. \eqref{GeeGeh_lr}. This situation is clearly seen in Fig. \ref{fig3}a in which weights of the two lowest-energy right eigenstates of the effective Hamiltonian, $|\psi_{1,2}|^2$, are displayed by red and blue solid curves, respectively.

			With the help of Eq. \eqref{GLARGB_fin}, the conductance contributions can be written as
			\begin{equation}
				\begin{split}
					G_{1L(R)}^{L} & =2G_{0}\frac{\frac{\Gamma_{1}^{2}}{4}\chi_{1(2),1(2N)}^{4}}{\left(eV\right)^2+4\beta_{1(2)}^2},\\
					G_{1L(R)}^{B} & =2G_{0}\frac{\underline{\chi}_{1A(2B)}^{T}
						Q_{\pm}\underline{\chi}_{1A(2B)}}{\left(eV\right)^2+4\beta_{1(2)}^2}\frac{\Gamma_{1}}{4}\chi_{1(2),1(2N)}^{2}.
				\end{split}
				\label{GLARGB_pt1}
			\end{equation}
			This result leads to $G_{1R(L)}^{B}=0$ if the bath is (anti-)Hermitian, i.e. when $\gamma_{1}+\tilde{\gamma}_{1}=\pm\eta_{s1}$. This effect is shown in Fig.\ref{fig3}f where $G_{1R}^{B}\left(V\right)$ is shown by the blue dash-dotted curve. We note that the small nonzero values of $G_{1R}^{B}\ll2G_{0}$ at $V\neq0$ and a weak asymmetry, $G_{1R}^{B}\left(V\right)\neq G_{1R}^{B}\left(-V\right)$, are due to a contribution to the $G_{1R}^{B}\left(V\right)$ from the higher-energy states which have been omitted in the analytical approach above.
			
			To advance further 
			we consider the symmetric point for which $\chi_{1(2),n}=\delta_{1(2N),n}$. Then, the eigenvalues become $j=L,R$
			\begin{equation}
				\beta_{1,2}^{(j)}=\frac{1}{4}\left[\left(\mu_{1,1(N)}\pm\nu_{1,1(N)}\right)^2+\Gamma_1\delta_{j,L(R)}\right] .
				\label{eigenval12_pt1_re_0}
			\end{equation}
			Taking it into account, we obtain the following results:
			\begin{equation}
				\begin{split}
					G_{1L(R)}^{L} & =\frac{2G_{0} \Gamma_1^2}{\left(2eV\right)^2+\left(\Gamma_1+\left(\mu_{1,1(N)}\pm\nu_{1,1(N)}\right)^2\right)^2},\\
					G_{1L(R)}^{B} & =\frac{2G_{0} \Gamma_1\left(\mu_{1,1(N)}\pm\nu_{1,1(N)}\right)^2}{\left(2eV\right)^2+\left(\Gamma_1+\left(\mu_{1,1(N)}\pm\nu_{1,1(N)}\right)^2\right )^2}. 
					\label{GLARGB_pt1_V}
				\end{split}
			\end{equation}
			The result \eqref{GLARGB_pt1_V} means the suppression of the Majorana zero-bias peak \cite{law-09,flensberg-10,ioselevich-13} by the dissipation. This phenomenon is clearly visible in Figs.\ref{fig3}b,c (see black solid curve). 
			
			As follows from the expressions \eqref{GLARGB_pt1_V} the quantization 
			of the right (left) zero-bias conductance can be achieved if the bath is (anti-)Hermitian, $\nu_{1,n}=\pm\mu_{1,n}$. For example, in Fig.\ref{fig3}f one can see that $G_{1R}(0)=G_{1R}^{L}(0)=2G_{0}$ for $\hat{L}_{1}=\hat{L}_{1}^{+}$. Qualitatively, this effect is explained by the presence of the zero kinetic mode (i.e. the state which is immune to the dissipation) in the absence of the lead, i.e. $\beta_{2}\left(\Gamma_{1}=0\right)=0$ in Eq. \eqref{eigenval12_pt1_re_0}, which dwells at the right edge. From the transport point of view, this case is identical to the dissipationless regime ($\hat{L}_{1}=0$) displayed by cyan dotted curve in Figs.\ref{fig3}. 
			
			On the other hand, if the lead is connected to the leftmost site of the chain ($j=L$) the transport is mediated by the only state with already finite lifetime $\sim 1/\beta_{1}\left(\Gamma_{1}=0\right)$ due to the bath. As a result, the quantization is destroyed as shown in Fig.\ref{fig3}e.
			
			To summarize, we considered a single-lead tunnel transport in the Kitaev chain with a weak 
				dissipative fields $\mu_{1,n},~\nu_{1,n}\in \mathbb{R}$, which are
				randomly varying in space. 
				If the bath is neither Hermitian nor anti-Hermitian (see the upper panel of Fig.\ref{fig3}) one cannot observe the zero-bias peak, $G_{1L(R)}\neq2G_0$. If $\hat{L}_{1}=\pm\hat{L}_{1}^{+}$ ($\nu_{1,n}=\mu_{1,n}$ in the lower panel of Fig.\ref{fig3}) the quantized conductance resonance at $V=0$ appears. According to Eq. \eqref{GLARGB_pt1_V} the dependence of the conductance on the connection point, i.e. $G_{1L}\left(V\right)\neq G_{1R}\left(V\right)$, is explained by the assumed spatial inhomogeneity of the dissipation.
			
			We note that if the bath fields are complex, $\mu_{1,n},~\nu_{1,n}\in \mathbb{C}$, then $|R_{1,2}\rangle$ become a linear superposition of $\underline{\chi}_{1,2}$ since $\left[\tilde{\mathcal{B}}\tilde{\mathcal{B}}^{T}\right]_{1,2}\neq0$. This case provides an additional opportunity to control the quantum transport by dissipation (
			the study of the corresponding effects is beyond the scope of the 
			present work). 
			
			Thus, the analysis performed above allows us to assert that the ubiquitous drawbacks of experiments aimed at detecting the Majorana mode, in particular, the asymmetry of the zero-bias peak and its non-quantized magnitude \cite{zhang-25}, can be caused by the interaction of the hybrid wire with some fermionic bath. 
			
			\begin{figure*}[!htb]
				\begin{center}
					\includegraphics[width=0.475\textwidth]{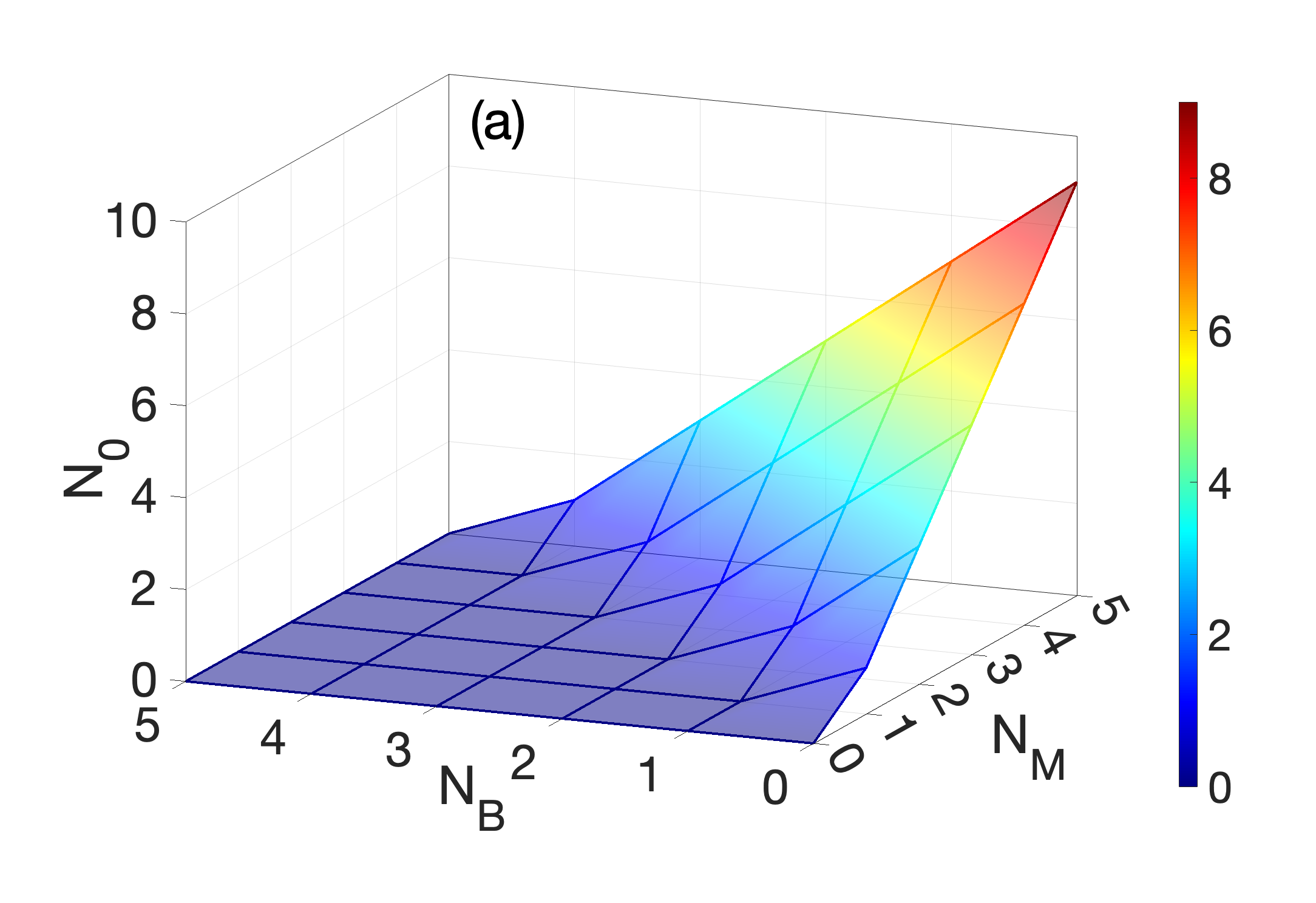}
					\includegraphics[width=0.475\textwidth]{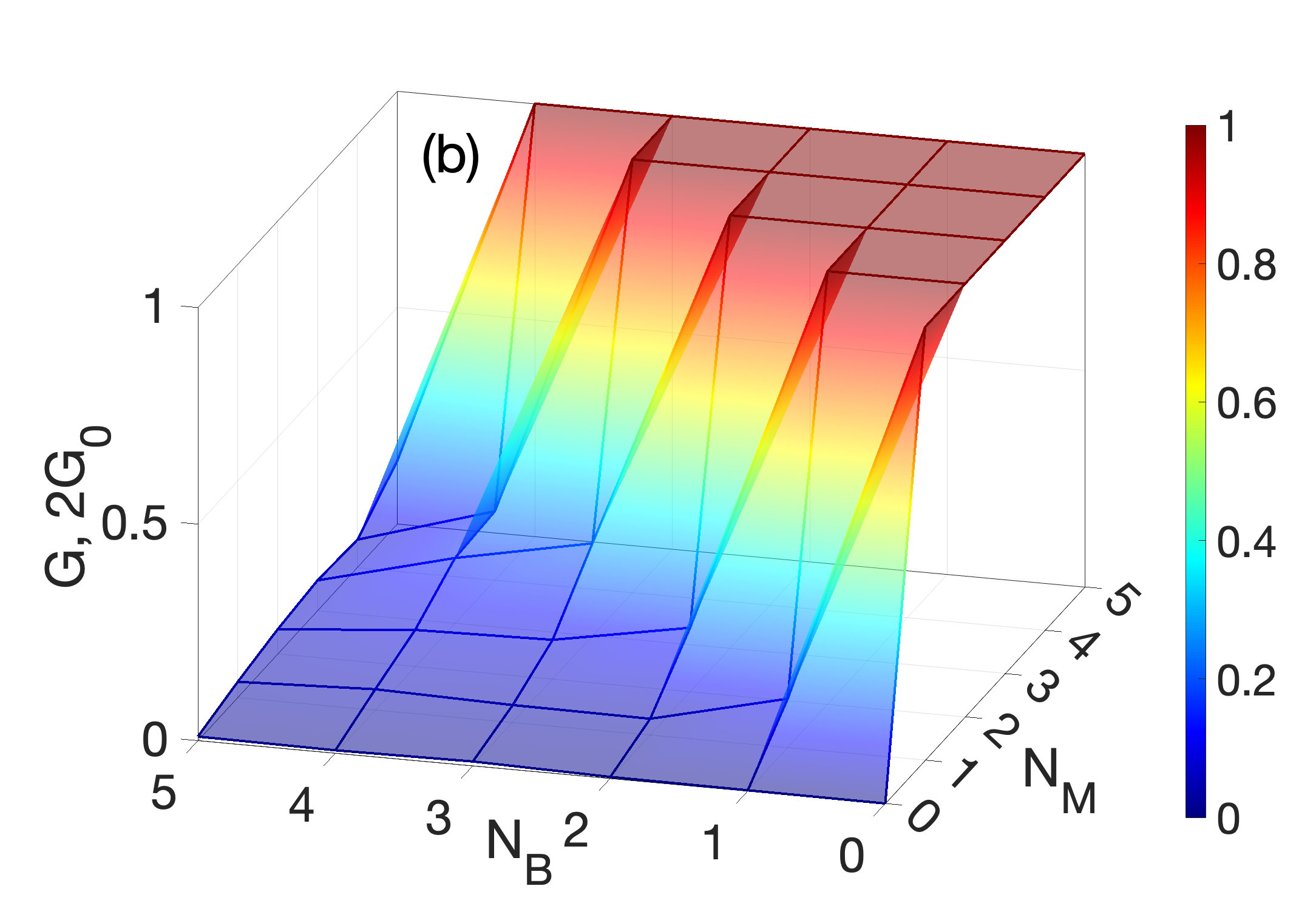}
					\includegraphics[width=0.475\textwidth]{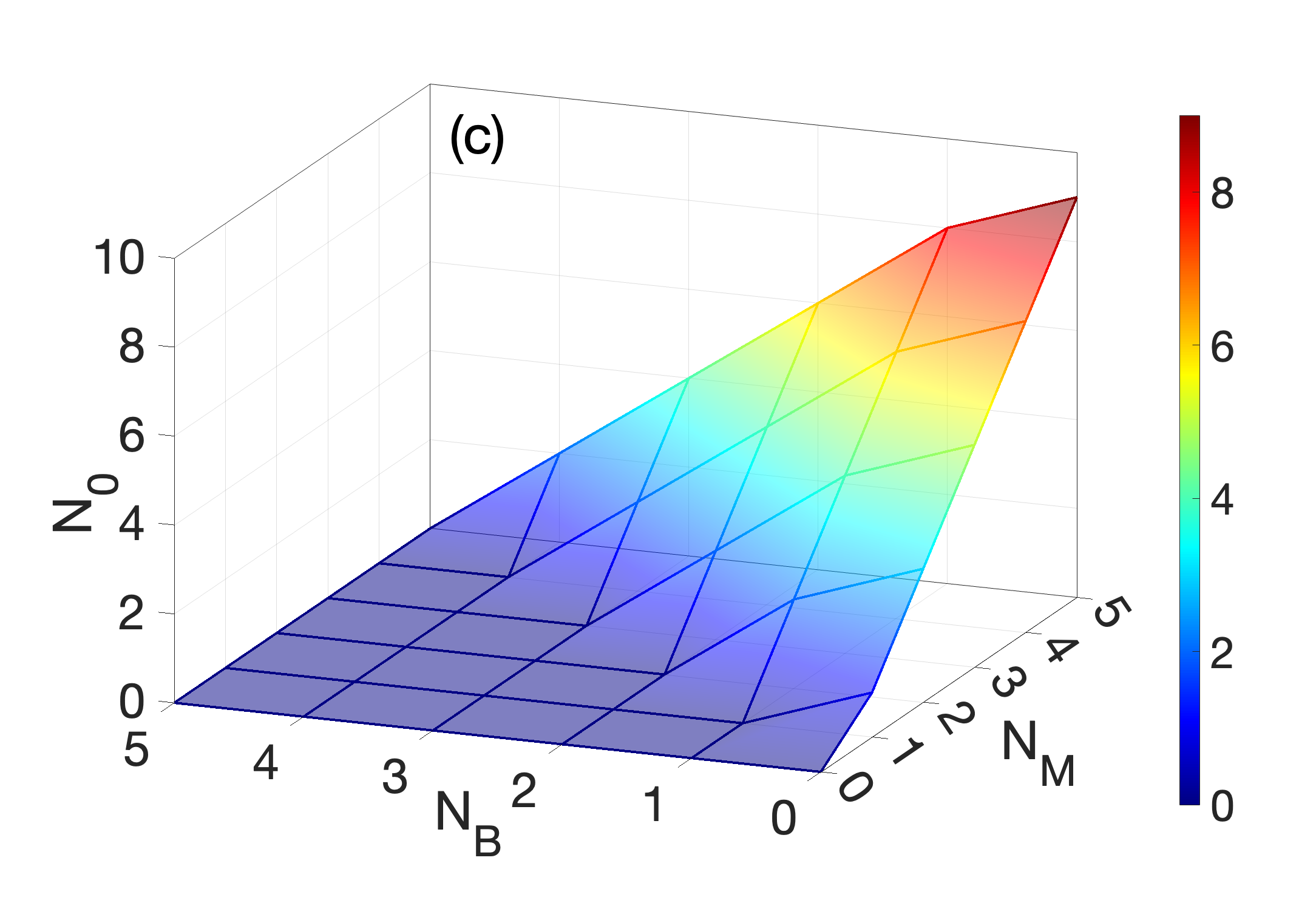}
					\includegraphics[width=0.475\textwidth]{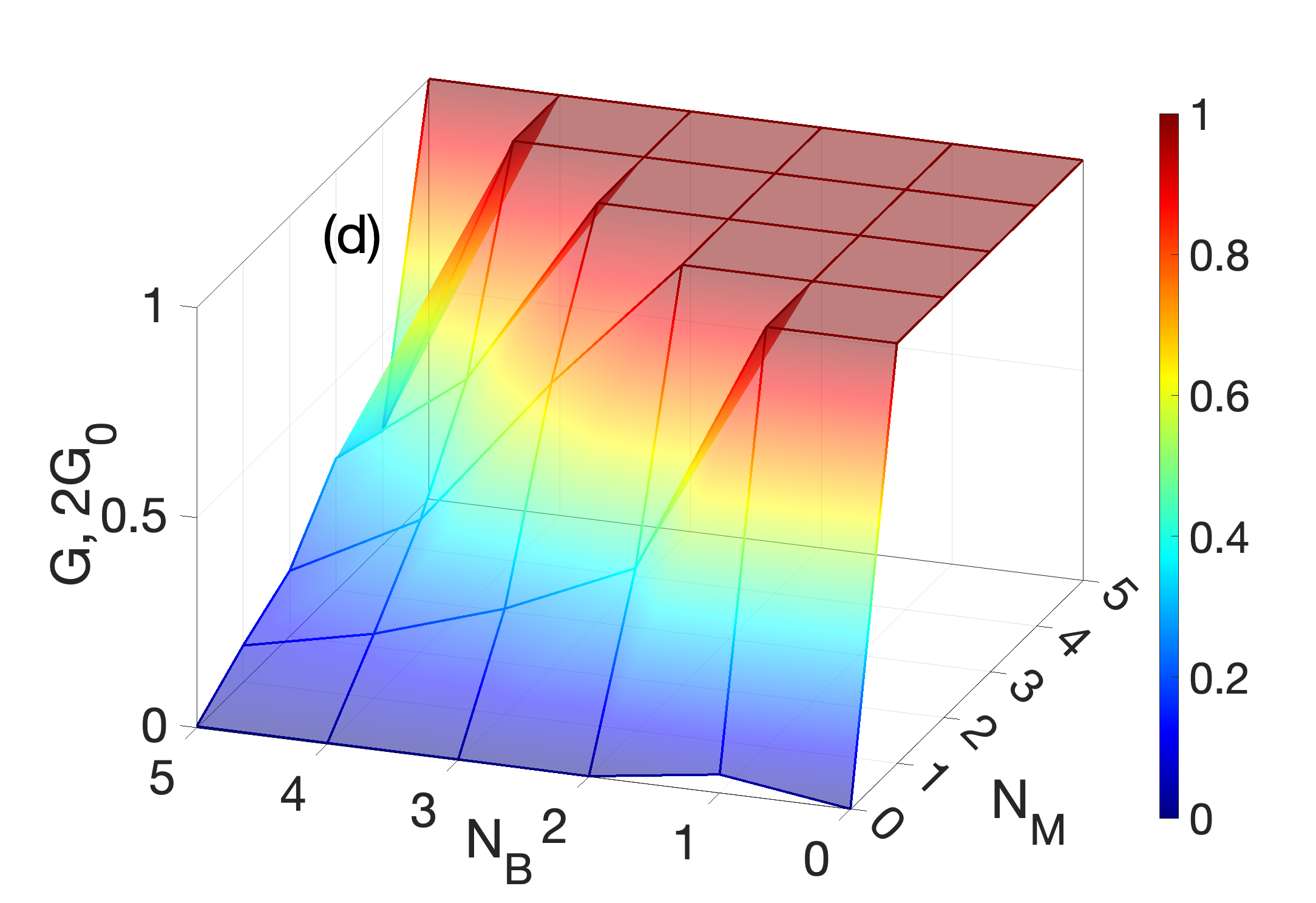}
    \caption{\label{MG_onecont} The effect of the NESS degeneracy on the local transport in the extended Kitaev model in the linear response and low-temperature regimes. The first column shows the dependence of the number of non-decaying zero-energy excitations ($N_0$) on the magnitude of the topological invariant of the isolated Kitaev chain ($N_{M}$), evaluated near the symmetric points, and on the number of baths ($N_B$) in the presence of the lead. The second column displays the corresponding dependence of the conductance $G$ of the lead connected to the leftmost site of the chain. (a,b) All bath Lindblad operators are non-Hermitian. (c,d) Only the Lindblad operator of the first bath is Hermitian. Parameters: $N=20$, $\mu=0.001$, $t=1$, $\Delta=-0.999$, $\Gamma_{1}=0.1$.}
				\end{center}
			\end{figure*}
			
			

			\subsection{Manifestation of NESS degeneracy in local transport measurements}\label{sec5.2}
			
			Finally, let us demonstrate how the NESS degeneracy affects the local transport in the case of $N_L = 1$, $N_{B}=0,\dots,5$. We consider the extended Kitaev model, where more than one pair of Majorana modes can emerge in the closed system.
			To ensure the presence of the required number of zero-energy excitations in the isolated system, we consider regions of the parameter space near the symmetric points of the extended Kitaev model with topological invariant $N_M$. In particular, in the Hamiltonian \eqref{Hs} we assume $\xi_{nn}=-\mu\approx0$ and $\xi_{nm}=t\delta_{m,n+N_M}=t\delta_{m,n-N_M}$, $\Delta_{nm}=\Delta\delta_{m,n+N_M}=-\Delta\delta_{m,n-N_M}$ with $t \approx -\Delta$ 
			\cite{degottardi-13}. In this regime, $N_M$ pairs of Majorana mode wavefunctions $\underline{\chi}_{2l-1}$ and $\underline{\chi}_{2l}$, where $l=1,\dots,N_M$, are localized at opposite ends of the chain.
			
			
			In Fig.~\ref{MG_onecont}a a numerical verification of the relation \eqref{M} for the case $N_L = 1$ with $t_{1n} = t_1 \delta_{n,N}$ is presented. The dependence of the number of zero-energy states $N_0$ on the value of the topological invariant $N_M$ and the number of the baths $N_B$ demonstrates exactly the same behavior as predicted by the theory of Ref. \cite{shustin-25}: in a generic case (where the bath fields on each site are random complex numbers, i.e., $\re\{\mu_{v,n}, \nu_{v,n}\}$, $\im\{\mu_{v,n}, \nu_{v,n}\} \in (0,1)$), increasing the number of baths by 1 reduces the NESS degeneracy by 2 when $N_M \geq N_B + N_L$. If $N_M < N_B + N_L$, the NESS is non-degenerate (see the triangular region with $N_0=0$ in Fig. \ref{MG_onecont}a).
			
			The two regions of degenerate and non-degenerate NESS in the dissipative system under consideration can be detected using tunnel spectroscopy by connecting a single-band normal paramagnetic lead to the edge of the wire. The corresponding conductance map in the linear response regime at low temperatures is shown in Fig.~\ref{MG_onecont}b. When the condition $N_{M} - N_{B} = 1$ holds, the system without the lead has two non-decaying zero-energy excitations. By connecting to the lead, as demonstrated in Appendix \ref{appB}, we remove one such state. The lifetime of this state is determined solely by the parameter $\Gamma_{1} = 2\pi\rho|t_{1}|^2$, whereas the Lamb shift is absent in the wide-band approximation. As a result, at small bias voltages, the quantum transport is governed exclusively by the resonant local Andreev reflection processes on the Majorana mode localized near the lead/chain boundary yielding $G = 2G_0$ \cite{law-09,flensberg-10,ioselevich-13}. The conductance quantization indicates the absence of the loss currents due to the dissipative processes.
			
			Thus, according to Eq. \eqref{M}, the boundary between degenerate and non-degenerate NESS in the space $\left(N_{M},~N_{B}\right)$ is the same for both $N_{L}=0$ and $N_{L}=1$. In the degenerate case, the conductance is quantized; otherwise, it is not. 
			We emphasize that the degree of degeneracy does not affect the quantization of the conductance.
			
			As noted in Sec. \ref{sec5.1} and Appendix \ref{appB}, the Hermiticity of the Lindblad operator corresponding to the bath increases the number of the non-decaying excitations in the open system. Figure \ref{MG_onecont}c shows the map of $M\left(N_{M},N_{B}\right)$ for the special case when the Lindblad operator of one of the baths is Hermitian, i.e., $\hat{L}_{1}=\hat{L}_{1}^{+}$; while for $v \neq 1$, $\hat{L}_{v} \neq \hat{L}_{v}^{+}$. In this case, in Eq. \eqref{M} we have $\dim\ker \tilde{\mathcal{B}}_B=1$, and the NESS is non-degenerate when $N_{M} \leq N_{B}$. In other words, the size of the region with $N_0=0$ does not change compared to the map in Fig. \ref{MG_onecont}a. However, the area of the quantized conductance plateau in Fig. \ref{MG_onecont}d increases, since at $N_{M} = N_{B}$ there now exists a zero-energy excitation immune to dissipation due to the first bath Hermiticity, which contributes to resonant transport processes.
			
			\section{Discussions and conclusions}\label{sec6}
			
			In this work the quantum transport in a superconducting device connected to an arbitrary number of normal single-band Fermi leads and subjected to dissipative processes is studied. The open nature of the device, due to interaction with an arbitrary number of fermionic baths, manifests in the fact that the system's density matrix (describing the device and the leads) exhibits non-unitary dynamics described by the GKSL equation with 
			quantum jump operators which are linear in creation and annihilation fermionic operators. Based on the quantum field-theoretical approach 
			on the Keldysh contour, an effective action is derived for the non-equilibrium open-type device. It is shown that the self-energy functions of the leads, unlike those of the baths, are nonlocal in time (even in the wide-band limit), i.e., they incorporate memory effects.
			
			Expressions describing the stationary charge and energy currents are obtained. In particular, we derived the extension of the Meir-Wingreen formula for description of the particle current in the superconducting device with dissipation. This current is determined by the device's retarded Green's function, which satisfies the Dyson equation incorporating the self-energies due to both contacts and baths. Without many-body effects the currents in the $i$-th lead can be formally represented as the sum of two contributions: the first is the sum of currents from all leads, and the second is the sum of currents from all baths. Each term in the first sum contains local components (local Andreev reflection) and nonlocal components (crossed Andreev reflection and elastic cotunneling). Each term in the second sum is the superposition of processes involving electron transfer from the individual bath to the $i$-th lead and back. 
			We have additionally assessed the loss current due to dissipation.
			
			In the linear-response regime and at low temperatures, a general form of the Onsager matrices characterizing thermoelectric effects is derived.
			Considering the presence of two contributions to the current in the $i$-th lead discussed above, one can distinguish two such matrices, each of which represents the sum of contributions from each lead and each bath, respectively.
			
			Essentially, although the dissipative quantum-transport equations are obtained for the arbitrary $\mu_{vn}$ and $\nu_{vn}$, the GKSL equation was derived in the assumption of a weak (tunnel) coupling between the system and the environment. This circumstance must be taken into account when considering specific models as Eq. \eqref{Lindblad_eq} may not be accurate if the smallness of $\mu_{vn}$ and $\nu_{vn}$ is 
				not enough
				\cite{levy-14,hofer-17,dechiara-18,reis-20,gauthameshwar-25}.
			
			
			
			The developed theory is applied to the widely known problem of charge transport in the Majorana modes in the Kitaev chain interacting with a single lead. It is shown that the dissipation causes the suppression of the conductance zero-bias peak. Additionally, we note that the stronger the coupling with the fermionic bath, the more pronounced the asymmetry of the Majorana resonance. These results may be useful for improving the data quality in the Majorana-state probing experiments.
			
			In the case of leads acting locally at single separate sites of the superconducting device, we demonstrated that under assumption of the wide-band approximation each lead reduces the NESS degeneracy by 1. These results are confirmed by numerical calculations for several cases of the extended Kitaev model at symmetric points with the single lead. Additionally, in the linear-response regime and at low temperatures, it is found that a (non-)degenerate NESS corresponds to the (non-)quantized conductance peak. 
			One can expect an observation of these effects in an array of superconducting quantum dots -- the system which has recently become popular in the light of the poor man's Majorana bound state formation \cite{tsintzis-22,dvir-23,tenhaaf-24}. In 
			such systems the ability to precisely control interactions between the dots has already enabled the realization of the symmetric point of the Kitaev model in the simplest double-dot case (also called the sweet spot). One of the intriguing questions here is the possibility to implement long-range interactions in such a system and to reach the multiply degenerate NESS in the corresponding sweet spots by means of the dissipative bath but not via the tuning of the inter-dot interactions (e.g. drawing on the idea of Ref. \cite{diehl-11} 
			where the Lindblad operator was chosen as the bulk Bogoliubov operator). If such a bath-engineered extended Kitaev model is possible, will it give rise to violation of the relations \eqref{M0} and \eqref{M}? 
			Another potential testbed to check the correlation between $N_{0,B,L,M}$ and $G\left(V=0\right)$ is a system of atomic fermions in 1D optical lattice coupled to baths and particle reservoirs \cite{corman-19}. 
			
			The obtained results open new opportunities for advancing the physics of non-equilibrium phenomena in low-dimensional normal and superconducting systems with dissipation.
			In particular, our theory allows to take into account the possible presence of fermion baths (affecting quantum transport) at the microscopic level. Their coupling with subgap states in superconducting device, e.g. in the experiments searching for the Majorana state in the hybrid nanowires, can have a significant impact on the measurements. From this point of view, one of further directions would be to compare the response of Majorana and Andreev bound states in the conductance of the dissipative wire. The latter are ubiquitous in the realistic samples due to disorder and non-uniform electrostatic potential \cite{prada-20,aksenov-23}. Study of thermoelectric effects in open-like transport scheme with topologically-ordered or strongly correlated media also seems promising. 
			
			Another direction is to study NESS properties and NESS design based on the obtained quantum kinetic equations \eqref{G+-ab}. For example, one can analyze how change in the NESS degeneracy affects the spatial behavior of the particle density and anomalous correlators. Finally, a fundamental problem that appears to be understudied is the coexistence of many-body effects and dissipation in the context of transport and NESS engineering. 
			
			\begin{acknowledgments}
				We are grateful to A. Melnikov, V. Khrapai and A. Kamenev for useful comments and discussions of the work. The work was carried out within the framework of the state assignment of the L.D. Landau Institute of Theoretical Physics of the Russian Academy of Sciences. The authors thank the BASIS Foundation for personal support. The authors acknowledge the hospitality during the international workshop "Landau Week 2025" where part of this work has been performed. 
			\end{acknowledgments}
			
			\appendix
			\section{\label{appA} Derivation of the Meir-Wingreen formula for a dissipative device}
			
			In this Appendix, we provide a detailed derivation of the Meir-Wingreen formula adapted for a superconducting device with dissipation.
			
			Integration of the lead states in the initial action \eqref{S} results in the following effective action,
			\begin{align}
				S_{\delta,eff} & =S_{0}+S_{\delta,a}\equiv \frac{1}{2}\int\limits_{-\infty}^{+\infty}dt\overline{\Psi}\left(i\partial_{t}\cdot I_{4N}-i\mathcal{L}_{s}\right)\Psi \notag \\
				& -\frac{1}{2}\sum\limits_{j}\iint\limits_{-\infty}^{+\infty}dt_{1}dt_{2}\overline{\Psi}\left(t_{1}\right)\left(S_{z}+ia_{j}\left(t_{1}\right)\xi_{jk}^{\zeta_{\delta}}S_{x}\right)\notag \\
				& \times \Sigma_{j}\left(t_{1}-t_{2}\right)\left(S_{z}-ia_{j}\left(t_{2}\right)\xi_{jk}^{\zeta_{\delta}}S_{x}\right)\Psi\left(t_{2}\right) ,\label{Seff}
			\end{align} 
			\noindent
			where $S_{0}=S_{\delta,eff}\left(a_{j}=0\right)$ denotes a bare action that includes couplings to the Fermi reservoirs and dissipative environments. In Eq. \eqref{Seff} we have introduced the self-energy matrix of the $j$-th lead, defined through its corresponding bare Green's function, i.e., $\Sigma_{j}\left(t_{1}-t_{2}\right)=\sum\limits_{k}T_{jk}^{+}G_{jk}\left(t_{1}-t_{2}\right)T_{jk}\equiv\sum\limits_{k}\Sigma_{jk}\left(t_{1}-t_{2}\right)$. In the frequency representation, the components of the latter in the Keldysh space are given as
			\begin{widetext}
				\begin{eqnarray}
					&&G_{jk}^{R}=\left( \begin{array}{*{20}{c}}
						\omega-\xi_{jk}+i\delta & 0 \\
						0 & \omega+\xi_{jk}+i\delta \end{array} \right)^{-1}\equiv \diag\{g_{je},g_{jh}\},~ G_{jk}^{A}=\left(G_{jk}^{R}\right)^{+},~\label{GkRA}\\
					&&G_{jk}^{K} = f_{j}\left(G_{jk}^{R}-G_{jk}^{A}\right) = -2\pi i\left( \begin{array}{*{20}{c}}
						\left(1-2n_{je}\right)\delta\left(\omega-\xi_{jk}\right) & 0 \\
						0 & \left(1-2n_{jh}\right)\delta\left(\omega+\xi_{jk}\right) \end{array} \right),\nonumber\\
					&& f_{j}=\diag\{f_{je},f_{jh}\},~f_{je(h)}=1-2n_{je(h)},~n_{je(h)}=\left(1+e^{\frac{\omega\mp eV_{j}}{T+T_{j}}}\right)^{-1}, \label{GkK}
				\end{eqnarray} 
			\end{widetext}
			\noindent
			where $T$ denotes the equilibrium temperature, $T_{j}$ stands for the deviation from $T$ in the $j$-th lead. 
			
			The generating function $Z_{\delta}\left[a_{1},...,a_{N_{L}}\right]=\int D\left[\overline{\psi}_{n}^{1,2},~\psi_{n}^{1,2},~\overline{\phi}_{jk}^{1,2},~\phi_{jk}^{1,2}\right] e^{iS_{\delta}}$ allows us to compute the average currents in $i$-th lead as
			\begin{widetext}
				\begin{eqnarray}\label{I2}
					I_{i,\delta} = \frac{ie^{\bar{\zeta}_{\delta}}}{2}\frac{\delta Z_{\delta}\left[a_{1},...,a_{N_{L}}\right]}{\delta a_{i}\left(t\right)}\Biggl|_{a_{i}=0}=\frac{ie^{\bar{\zeta}_{\delta}}}{2}\sum\limits_{nk}\left\langle\xi_{ik}^{\zeta_{\delta}} \left[t_{in}\left(\psi_{n}^{1}\overline{\phi}_{ik}^{2}+\psi_{n}^{2}\overline{\phi}_{ik}^{1}\right)-t_{in}^{*}\left(\phi_{ik}^{1}\overline{\psi}_{n}^{2}+\phi_{ik}^{2}\overline{\psi}_{n}^{1}\right)\right]\right\rangle,
				\end{eqnarray}
			\end{widetext}
			\noindent
			where $\bar{\zeta}_{\delta}=1,0$ if $\delta=c,\varepsilon$. Inserting $S_{\delta,eff}$ into the generating function $Z_{\delta}\left[a_{1},...,a_{N_{L}}\right]$, 
			taking the functional derivative in Eq. \eqref{I2} results in
				\begin{widetext}
					\begin{equation}
						I_{i,\delta}  = -i\frac{e^{\bar{\zeta}_{\delta}}}{4}\int D\left[\overline{\psi}_{n}^{1,2},~\psi_{n}^{1,2}\right] e^{iS_{0}}  \sum\limits_{k}\int\limits_{-\infty}^{+\infty}dt' \xi_{ik}^{\zeta_{\delta}} \left[\overline{\Psi}\left(t\right)S_{x}\Sigma_{ik}\left(t-t'\right)S_{z}\Psi\left(t'\right)
						-\overline{\Psi}\left(t'\right)S_{z}\Sigma_{ik}\left(t'-t\right)S_{x}\Psi\left(t\right)\right].
						\label{I3a}
					\end{equation}
				\end{widetext}
			
			Using relations $\overline{\Psi}S_{x}\Sigma_{i}S_{z}\Psi=\Tr\left\{S_{x}\Sigma_{i}S_{z}\Psi\overline{\Psi}\right\}$ and $\overline{\Psi}S_{z}\Sigma_{i}S_{x}\Psi=\Tr\left\{S_{z}\Sigma_{i}S_{x}\Psi\overline{\Psi}\right\}$, one can obtain after the Fourier transform: 
			\begin{equation}
				\begin{split}
					&I_{i,\delta}  = \frac{e^{\bar{\zeta}_{\delta}}}{4}\int\limits_{-\infty}^{+\infty}\frac{d\omega}{2\pi}\omega^{\zeta_{\delta}}\Tr\left\{\left[S_{z}^{\bar{\zeta}_{\delta}}\Sigma_{i},S_{x}\right]G_{s}\right\} \\
					& =\frac{ie^{\bar{\zeta}_{\delta}}}{4}\int\limits_{-\infty}^{+\infty}\frac{d\omega}{2\pi} \omega^{\zeta_{\delta}}\Tr\left\{\tau_{z}^{\bar{\zeta}_{\delta}}\Gamma_{i}\left[G_{s}^{K}{-}F_{i}
					\left(G_{s}^{R}{-}G_{s}^{A}\right)
					\right]\right\}. 
					\label{I4}
				\end{split}
			\end{equation}
			In the first equality in Eq. \eqref{I4} the system's Green's function is introduced $G_{s}=-i\left\langle\Psi\overline{\Psi}\right\rangle$, where averaging is performed with respect to the bare action $S_{0}$.
			Accordingly, the second equality in \eqref{I4} employs the blocks of this matrix in the Keldysh space, $G_{s}^{R,A,K}$. Using Eqs. \eqref{GkRA} and \eqref{GkK}, the matrices of the distribution functions and the level broadenings of the system due to coupling with the 
			$i$-th lead are expressed in the Bogoliubov-de Gennes form as
			\begin{eqnarray}
				&&F_{i}=\diag\{F_{ie},~F_{ih}\}=f_{i} \otimes I_{N},~\label{Fi}\\
				&&\Gamma_{i}=\diag\{\Gamma_{ie},~\Gamma_{ih}\},~\Gamma_{ih}\left(\omega\right)=\Gamma_{ie}^{T}\left(-\omega\right),\label{Gammaieh}\\
				&&\Gamma_{ie}=i\left(\Sigma_{ie}^{R}-\Sigma_{ie}^{A}\right)=2\pi \sum\limits_{k}\underline{t}_{ik}^{*}\underline{t}_{ik}^{T}\delta\left(\omega-\xi_{ik}\right),\label{Gammaie}\\
				&&\Gamma_{ih}=i\left(\Sigma_{ih}^{R}-\Sigma_{ih}^{A}\right)=2\pi \sum\limits_{k}\underline{t}_{ik}\underline{t}_{ik}^{+}\delta\left(\omega+\xi_{ik}\right).~~~\label{Gammaih}
			\end{eqnarray}
			
			The expression \eqref{I4} with $\zeta_{\delta=c}=0$ and $N_{L}=2$ generalizes the Meir-Wingreen formula \cite{meir-92} to the case of a superconducting device with dissipation. It describes the charge current in the $i$-th lead which is connected to a region where both dissipation processes and many-particle interactions may be present. Let us now examine how Eq. \eqref{I4} transforms when the many-body effects are absent, i.e., in the case of quadratic Liouvillians.
			
			In this situation the Fourier transform of the bare effective action 
			$S_{0}$ yields the inverse Green's function $G_{s}^{-1}\left(\omega\right)$,
			\begin{widetext}
				\begin{eqnarray}\label{S0eff2}
					S_{0} = \frac{1}{2}\int\limits_{-\infty}^{+\infty}\frac{d\omega}{2\pi}
					\left(\overline{\Psi}^{1}~\Psi^{2}~\overline{\Psi}^{2}~\Psi^{1}\right)
					\left( \begin{array}{*{20}{c}}
						\omega I_{2N}-\sum\limits_{j}\Sigma_{j}^{R}-i\mathcal{L}_{s}^{R} & -\sum\limits_{j}\Sigma_{j}^{K}-i\mathcal{L}_{s}^{K} \\
						0 & \omega I_{2N}-\sum\limits_{j}\Sigma_{j}^{A}-i\mathcal{L}_{s}^{A} \end{array} \right)
					\left( \begin{array}{*{04}{c}}
						\Psi^{1} \\ \overline{\Psi}^{2} \\ \Psi^{2} \\ \overline{\Psi}^{1} \end{array} \right),
				\end{eqnarray}
			\end{widetext}
			\noindent
			where 
			\begin{eqnarray}\label{SigmaRAK}
				\Sigma_{j}^{R,A} = \Lambda_{j}\mp\frac{i}{2}\Gamma_{j},~\Sigma_{j}^{K} = -i\Gamma_{j}F_{j}.
			\end{eqnarray}
			The self-energy \eqref{SigmaRAK} 
			involves the energy level shifts of the system induced by its interaction with the $j$-th lead,
			\begin{gather}
				\Lambda_{j}=\diag\{\Lambda_{je}, \quad \Lambda_{jh}\}, \quad\Lambda_{jh}\left(\omega\right)=-\Lambda_{je}^{T}\left(-\omega\right), \notag \\
				\Lambda_{je}=\sum\limits_{k}\textrm{P.v.}\frac{\underline{t}_{jk}^{*}\underline{t}_{jk}^{T}}{\omega-\xi_{jk}}, \qquad \Lambda_{jh}=\sum\limits_{k}\textrm{P.v.}\frac{\underline{t}
					_{jk}\underline{t}_{jk}^{+}}{\omega+\xi_{jk}} .\label{Lambda}
			\end{gather}
			
			Building on the results of Ref. \cite{thompson-23}, which similarly analyzed linear Lindblad operators \eqref{Lvn}, we finally obtain the following diagonal, $\left(G_{s}^{-1}\right)^{R,A}$, and off-diagonal, $\left(G_{s}^{-1}\right)^{K}$, blocks of the inverse Green's function of the device:
			\begin{eqnarray}\label{iG_app}
				&&\left(G_{s}^{-1}\right)^{R,A} = \omega I_{2N}-h_{s}-\sum\limits_{j}\Lambda_{j}\pm\frac{i}{2}\Bigl(\sum\limits_{j}\Gamma_{j}+\sum\limits_{v}Q_{v}\Bigr),\nonumber\\
				&&\left(G_{s}^{-1}\right)^{K} = i\Bigl(\sum\limits_{j}\Gamma_{j}F_{j}+\sum\limits_{v}D_{v}\Bigr).
			\end{eqnarray}
			The explicit form of the matrices $Q_{v}$ and $D_{v}$, which characterize the influence of dissipative processes on the superconducting system, was previously derived in Ref. \cite{thompson-23},
			\begin{equation}
				Q_{v}{=}\begin{pmatrix}
					\gamma_{v}{+}\tilde{\gamma}_{v} & \eta_{sv}^{+} \\
					\eta_{sv} & \gamma_{v}^{T}{+}\tilde{\gamma}_{v}^{T}
				\end{pmatrix}, \quad 
				D_{v}{=}\begin{pmatrix}
					\gamma_{v}{-}\tilde{\gamma}_{v} & \eta_{av}^{+} \\
					\eta_{av} & \tilde{\gamma}_{v}^{T}{-}\gamma_{v}^{T}
				\end{pmatrix} ,
				\label{Qf}
			\end{equation}
			Here we introduced
			\begin{gather}
				\gamma_{v}=\underline{\mu}_{v}^{*}\underline{\mu}_{v}^{T},~\tilde{\gamma}_{v}=\underline{\nu}_{v}\underline{\nu}_{v}^{+},~\eta_{v}=\underline{\nu}_{v}^{*}\underline{\mu}_{v}^{T},~\eta_{vs(a)}=\eta_{v}\pm\eta_{v}^{T},\nonumber\\
				\gamma_{v}=\gamma_{v}^{+},~\tilde{\gamma}_{v}=\tilde{\gamma}_{v}^{+},~\eta_{s(a)v}=
				\pm\eta_{s(a)v}^{T} ,
			\end{gather} 
			where $\underline{\mu}_{v}=\left(\mu_{v1},...,\mu_{vN}\right)^{T}$, $\underline{\nu}_{v}=\left(\nu_{v1},...,\nu_{vN}\right)^{T}$. 
			
			Based on the structure of the matrix $G_{s}^{-1}$, we can arrive at the following relation: $$G_{s}^{R}\Bigl(\sum\limits_{j}\Gamma_{j}+\sum\limits_{v}Q_{v}\Bigr)G_{s}^{A}=i\left(G_{s}^{R}-G_{s}^{A}\right).$$
			As a result, taking into account the Keldysh equation, $G_{s}^{K}=-G_{s}^{R}\left(G_{s}^{-1}\right)^{K}G_{s}^{A}$, the stationary average currents in the $i$-th lead are given by
			\begin{gather}
				I_{i,\delta} {=} \frac{e^{\bar{\zeta}_{\delta}}}{4}\int\limits_{-\infty}^{+\infty}\frac{d\omega}{2\pi} \omega^{\zeta_{\delta}} \Tr\Bigl\{\tau_{z}^{\bar{\zeta}_{\delta}}\Gamma_{i}\Bigl[G_{s}^{R}\Bigl(\sum\limits_{j}\Gamma_{j}F_{j}{+}\sum\limits_{v}D_{v}\Bigr)G_{s}^{A}\notag \\ {-} F_{i}G_{s}^{R}\Bigl(\sum\limits_{j}\Gamma_{j}{+}\sum\limits_{v}Q_{v}\Bigr)G_{s}^{A}\Bigr]\Bigr\},
				\label{Ifin1}
			\end{gather}
			where the retarded Green's function obeys the modified Dyson equation,
			\begin{eqnarray}\label{Dysoneq}
				&&G_{s}^{R} = \Bigl(\omega I_{2N}-h_{s}-\sum\limits_{j}\Lambda_{j}+\frac{i}{2}\Bigl[\sum\limits_{j}\Gamma_{j}+\sum\limits_{v}Q_{v}\Bigr]\Bigr)^{-1},\nonumber\\
				&&G_{s}^{A} = \left(G_{s}^{R}\right)^{+}.
			\end{eqnarray}
			
			From Eq. \eqref{iG_app}, it is evident that the dissipation causes the appearance of new terms, $\sim\sum\limits_{v}D_{v}$ and $\sim\sum\limits_{v}Q_{v}$, in the 
			self-energy functions, respectively. As a result, additional components arise in the current \eqref{Ifin1}. Taking this into account, we will distinguish the contributions to the current in the $i$-th lead from the Fermi reservoirs and from the baths (while keeping in mind that both contributions themselves depend on the characteristics of the leads and the baths),
			\begin{eqnarray}\label{ILB_app}
				I_{i,\delta}=I_{i,\delta}^{L}+I_{i,\delta}^{B}. 
			\end{eqnarray}
			Then, taking into account the block structure of the matrices appearing in Eq. \eqref{Ifin1} in the electron-hole space, for example,
			$G_{s}^{R}=\left( \begin{array}{*{20}{c}}
				G_{ee} & G_{eh} \\
				G_{he} & G_{hh} \\
			\end{array} \right)$, it is not difficult to show that the contribution to the current due to the leads is equal to
			\begin{align}\label{IfinC_app}
				I_{i,\delta}^{L} & = e^{\bar{\zeta}_{\delta}}\sum\limits_{j}\int\limits_{-\infty}^{+\infty}\frac{d\omega}{2\pi} \omega^{\zeta_{\delta}}\Biggl [\Tr\left(\Gamma_{ie}G_{ee}\Gamma_{je}G_{ee}^{+}\right)\left(n_{ie}-n_{je}\right) \notag \\
				&+\Tr\left(\Gamma_{ie}G_{eh}\Gamma_{jh}G_{eh}^{+}\right)\left(n_{ie}-n_{jh}\right)  \Biggr ]  \notag\\
				&\equiv e^{\bar{\zeta}_{\delta}}\sum\limits_{j}\sum\limits_{\alpha=e,h}\int\limits_{-\infty}^{+\infty}\frac{d\omega}{2\pi} \omega^{\zeta_{\delta}}T_{e\alpha}^{(i,j)}\left(n_{ie}-n_{j\alpha}\right).
			\end{align}
			In the derivation of Eq. \eqref{IfinC_app}, the symmetry properties of the Green's function, $G_{ee}\left(\omega\right)=-G_{hh}^{*}\left(-\omega\right)$, $G_{eh}\left(\omega\right)=-G_{he}^{*}\left(-\omega\right)$, and the matrices of the broadening, Eq. \eqref{Gammaih}, were used. In the first equality, the first term describes non-local electron propagation (elastic cotunneling), $I_{i,\delta}^{EC}$, while the second term corresponds to the local ($i=j$), $I_{i,\delta}^{LAR}$, and non-local ($i \neq j$), $I_{i,\delta}^{CAR}$, Andreev reflection processes. In the second equality, effective transmission coefficients are formally introduced, characterizing the probabilities of the described transport processes.
			
			The contribution to the current in the $i$-th lead due to the dissipative processes is given as
			\begin{widetext}
				\begin{align}
					I_{i,\delta}^{B} & = \frac{e^{\bar{\zeta}_{\delta}}}{2}\sum\limits_{vk}\int\limits_{-\infty}^{+\infty}d\omega \omega^{\zeta_{\delta}}\left\{\left[\langle\,\underline{t}_{ik}^{T}\underrightarrow{b}_{ev}\,|\,\underline{t}_{ik}^{T}\underrightarrow{b}_{ev}\,\rangle n_{ie} - \langle\,\underline{t}_{ik}^{T}\underleftarrow{b}_{ev}\,|\,\underline{t}_{ik}^{T}\underleftarrow{b}_{ev}\,\rangle \left(1-n_{ie}\right)\right]\delta\left(\omega-\xi_{ik}\right)\right.\nonumber\\
					&- \left. \left(-1\right)^{\zeta_{\delta}}\left[\langle\,\underline{t}_{ik}^{+}\underrightarrow{b}_{hv}\,|\,\underline{t}_{ik}^{+}\underrightarrow{b}_{hv}\,\rangle n_{ih} - \langle\,\underline{t}_{ik}^{+}\underleftarrow{b}_{hv}\,|\,\underline{t}_{ik}^{+}\underleftarrow{b}_{hv}\,\rangle \left(1-n_{ih}\right)\right]\delta\left(\omega+\xi_{ik}\right) \right\}\nonumber\\
					&\equiv e^{\bar{\zeta}_{\delta}}\sum\limits_{v}\int\limits_{-\infty}^{+\infty}\frac{d\omega}{2\pi} \omega^{\zeta_{\delta}}\left[\underrightarrow{T}_{e}^{(i,v)}n_{ie}-\underleftarrow{T}_{e}^{(i,v)}\left(1-n_{ie}\right)\right],
					\label{IfinB_app}
				\end{align}
			\end{widetext}
			\noindent
			where
			\begin{align}
				\underrightarrow{b}_{ev}{=}G_{ee}\,\underline{\mu}_{v}^{*}{+}G_{eh}\,\underline{\nu}_{v}^{*}, &\quad \underleftarrow{b}_{ev}{=}G_{ee}\,\underline{\nu}_{v}{+}G_{eh}\,\underline{\mu}_{v},\notag \\
				\underrightarrow{b}_{hv}{=}G_{he}\,\underline{\mu}_{v}^{*}{+}G_{hh}\,\underline{\nu}_{v}^{*}, & \quad \underleftarrow{b}_{hv}{=}G_{he}\,\underline{\nu}_{v}{+}G_{hh}\,\underline{\mu}_{v}. \label{beh_app}
			\end{align}
			The rates of the transport processes for electrons and holes in Eq. \eqref{IfinB_app} are understood in the context of the Hilbert-Schmidt scalar product, i.e.,
			\begin{align}\label{Teh_rates_app}
				\underrightarrow{T}_{e(h)}^{(i,v)}&=2\pi\sum\limits_{k} \langle\,\underline{t}_{ik}^{T}\underrightarrow{b}_{e(h)v}\,|\,\underline{t}_{ik}^{T}\underrightarrow{b}_{e(h)v}\,\rangle\delta\left(\omega\mp\xi_{ik}\right)\notag\\
				&=\Tr\left\{\Gamma_{ie(h)}\underrightarrow{b}_{e(h)v}\underrightarrow{b}_{e(h)v}^{+}\right\},\\
				\underleftarrow{T}_{e(h)}^{(i,v)}&=2\pi\sum\limits_{k}\langle\,\underline{t}_{ik}^{T}\underleftarrow{b}_{e(h)v}\,|\,\underline{t}_{ik}^{T}\underleftarrow{b}_{e(h)v}\,\rangle\delta\left(\omega\mp\xi_{ik}\right)\nonumber\\
				&=\Tr\left\{\Gamma_{ie(h)}\underleftarrow{b}_{e(h)v}\underleftarrow{b}_{e(h)v}^{+}\right\}.
			\end{align}
			Again taking into account the symmetry properties of the Green's functions and the broadening matrices, we obtain the relations $\underrightarrow{T}_{e}^{(i,v)}\left(\omega\right)=\underleftarrow{T}_{h}^{(i,v)}\left(-\omega\right)$ and $\underleftarrow{T}_{e}^{(i,v)}\left(\omega\right)=\underrightarrow{T}_{h}^{(i,v)}\left(-\omega\right)$, that were used in the formulation of the second equality in \eqref{IfinB_app}.
			
			\section{\label{appB} The effect of the leads on the spectrum of the dissipative superconducting system for $N_{B}+N_{L}\leq2$}
			
			In this Appendix, we examine the influence of leads on the spectrum of the open one-dimensional superconducting system consisting of $N$ sites, for several representative cases with $N_{B}+N_{L}\leq2$. This choice is motivated by the fact that in the tunneling spectroscopy of one-dimensional systems it is common to use one or two normal leads (probes) \cite{yu-21,aghaee-23}.
			
			Let us assume that for $N_{B,L}=0$ the isolated wire described by the Hamiltonian \eqref{Hs} realizes a topological phase with a topological invariant equal to $\pm N_{M}$. In other words, for open boundary conditions, there exist $2N_{M}$ Bogoliubov excitations with zero energy, $\varepsilon_{m}=0$, $m=1,...,2N_{M}$ (the factor of $2$ reflects the particle-hole symmetry of the model). 
			
			The wave function of any Bogoliubov excitation with energy $\varepsilon_{l}$ ($l=1,...,2N$) can be represented as a superposition of a pair of Majorana-mode wave functions $\underline{\chi}_{2l-1}$, $\underline{\chi}_{2l}$ \cite{valkov-22}. The spectral properties of the non-Hermitian Hamiltonian matrix $\tilde{X}$ \eqref{Xmod} are analyzed in the basis $\underline{\chi}_{a}$, $a=1,...,2N$. Hence, for a single lead connected to the edge site, i.e., $\underline{t}_{1}=t_{1,n}\delta_{1,n}$, and in the absence of baths $N_{B}=0$, the constant term of the characteristic equation $\tilde{\mathcal{P}}_{1}(\beta)\equiv\det\left(\tilde{X}-\beta I_{2N}\right)=0$ is equal to (we took into account Eq. \eqref{Ups})
			\begin{align}
				\tilde{\mathcal{P}}_{1}(0) & = 
				\prod\limits_{l=1}^{N}\varepsilon_{l}^{2}+\left[\sum\limits_{l=1}^{N}\det\mathcal{C}_l^{(1)}\prod\limits_{s \neq l}\varepsilon_{s}\right]^{2}\notag \\
				&=\prod\limits_{l=1}^{N}\varepsilon_{l}^{2}+\left[\sum\limits_{l=1}^{N}|t_{1,1}|^2\chi_{2l-1,1}\chi_{2l,1}\prod\limits_{s \neq l}\varepsilon_{s}\right]^{2},\label{Pbet01}
			\end{align}
			where $ \mathcal{C}_l^{(1)} $ is the matrix describing the hybridization of the lead fields $ \underline{t}_{1}^{r,i} $ with the wave functions of the Majorana modes associated with the $l$-th Bogoliubov excitation,
			\begin{eqnarray}\label{C_matr1}
				\mathcal{C}_l^{(1)} = \left( \begin{array}{*{20}{c}}
					\,\underline{\chi}_{2l-1}^{T}\underline{t}_{1}^{r}\, & ~~~\,\underline{\chi}_{2l-1}^{T}\underline{t}_{1}^{i}\, \\
					~~\,\underline{\chi}_{2l}^{T}\underline{t}_{1}^{r}\, & ~~~~~\,\underline{\chi}_{2l}^{T}\underline{t}_{1}^{i}\, \\
				\end{array} \right) \in \mathbb{R}^{2\times 2}. 
			\end{eqnarray}
			
			Then, for $N_{M} < N_{B} + N_{L} = 1$ there are no zeros, $N_0=0$, which also follows from Eq. \eqref{Btilde} since $\tilde{\mathcal{B}} = 0$. If $N_{M} = 1$, then according to Eq. \eqref{M} the single zero is possible because $\mathcal{B}_B = 0$ and $r = 0$, that is accompanied by the disappearance of the constant term,
			\begin{eqnarray}\label{Pbet01_NBDI1}
				\tilde{\mathcal{P}}_{1}(\beta=0) &=& 
				|t_{1,1}|^4\chi_{1,1}^2\chi_{2,1}^2\prod\limits_{s \neq 1}\varepsilon_{s}^2.
			\end{eqnarray}
			The latter is guaranteed precisely in the non-trivial phase, since $\chi_{1,1} \to 0$ ($\chi_{2,1} \to 1$) or $\chi_{2,1} \to 0$ ($\chi_{1,1} \to 1$) as $N \to \infty$. Thus, in the conventional Kitaev model with the nearest-neighbor interactions, non-equilibrium induced by the presence of lead connected to only one Majorana mode of the excitation with $\varepsilon_{1} = 0$ leaves the ground state doubly degenerate. This fact is a consequence of the topological protection of the Majorana state. It is known that lifting this degeneracy in the absence of dissipative processes induces destructive interference processes in the quantum transport \cite{flensberg-10,vuik-19}.
			
			For $N_{M} > 1$ and $N \to \infty$, the rectangular matrix $\tilde{\mathcal{B}}$ \eqref{Btilde} has the following structure
			\begin{eqnarray}\label{Bdef_1C}
				\tilde{\mathcal{B}} = \left( \begin{array}{*{20}{c}}
					t_{1,1}^{r}\chi_{1,1} & t_{1,1}^{i}\chi_{1,1} \\
					0 & 0 \\
					\vdots & \vdots \\
					t_{1,1}^{r}\chi_{2l-1,1} & t_{1,1}^{i}\chi_{2l-1,1} \\
					0 & 0 \\
					\vdots & \vdots \\
					t_{1,1}^{r}\chi_{2N_{M}-1,1} & t_{1,1}^{i}\chi_{2N_{M}-1,1} \\
					0 & 0 
				\end{array} \right),
			\end{eqnarray}
			where the vanishing of the even-numbered rows is due to the spatial separation of the Majorana modes forming the $N_{M}$ electron-like edge states (upon shifting the phase of the superconducting order parameter by $\pi$, the parity of the zero rows changes). Consequently, $2N_{M}-1$ zero-energy excitations remain in the open system.
			
			The constant term in the general case of two pairs of arbitrary fields (i.e., $N_{B}+N_{L}=2$, $N_{B,L}\geq0$) and the arbitrary number of zero modes in the closed system takes the form:
			\begin{widetext}
				\begin{gather}
					\tilde{\mathcal{P}}_{2}(0) = 
					\prod\limits_{l=1}^{N}\varepsilon_{l}^{2}+\frac{1}{2}\sum\limits_{n,m=1}^{4}\left[\sum\limits_{l=1}^{N}\det\mathcal{C}_l^{(n,m)}\prod\limits_{s \neq l}\varepsilon_{s}\right]^{2}
					+\left[\sum\limits_{l_{1}\neq l_{2}=1}^{N}\sum\limits_{n,m,k,p=1}^{4}\frac{1}{4}\varepsilon_{nmkp}\det\mathcal{C}_{l_{1}}^{(n,m)} \det\mathcal{C}_{l_{2}}^{(k,p)}\prod\limits_{s \neq l_{1},l_{2}}\varepsilon_{s}\right]^{2},\label{Pbet02}
				\end{gather}
			\end{widetext}
			where $\varepsilon_{nmkp}$ is the Levi-Civita symbol. The matrices $\mathcal{C}_l^{(n,m)}$ are a generalization of the definition in Eq. \eqref{C_matr1}. They describe the hybridization between the Majorana modes of the excitation with energy $\varepsilon_{l}$ and arbitrary dissipative fields $\underline{\tilde{l}}_{n,m}$, i.e., $\underline{\tilde{l}}_{n,m} = \underline{l}_{n,m}^{r(i)}, \underline{t}_{n,m}^{r(i)}$,
			\begin{eqnarray}\label{C_matr2}
				\mathcal{C}_l^{(n,m)} = \left( \begin{array}{*{20}{c}}
					\,\underline{\chi}_{2l-1}^{T}\underline{\tilde{l}}_{n}\, & ~~~\,\underline{\chi}_{2l-1}^{T}\underline{\tilde{l}}_{m}\, \\
					~~\,\underline{\chi}_{2l}^{T}\underline{\tilde{l}}_{n}\, & ~~~~~\,\underline{\chi}_{2l}^{T}\underline{\tilde{l}}_{m}\, \\
				\end{array} \right) \in \mathbb{R}^{2\times 2}. 
			\end{eqnarray}
			It follows again from Eq. \eqref{Pbet02} that NESS degeneracy occurs only when $N_{M} > 0$.
			
			Let us now consider the case of two normal leads (i.e. $N_{B} = 0$) coupled to the opposite ends of the wire, $\underline{t}_{1}=t_{1,n}\delta_{1,n}$ and $\underline{t}_{2}=t_{2,n}\delta_{N,n}$. In turn, it means that
			\begin{eqnarray}
				&&\underline{\tilde{l}}_{1(2)}=\left(t_{1,1}^{r(i)},...,0,\mp t_{1,1}^{i(r)},0,...,0\right)^{T}, \notag \\
				&&\underline{\tilde{l}}_{3(4)}=\left(0,...,0,t_{2,N}^{r(i)},0,...,0,\mp t_{2,N}^{i(r)}\right)^{T}.\label{l1234}
			\end{eqnarray}
			Then, for any spatially non-local Bogoliubov excitation with $\varepsilon_{l} = 0$, we have:
			\begin{eqnarray}
				&&\det\mathcal{C}_l^{(1,2)}=\det\mathcal{C}_l^{(3,4)}=0, \notag \\
				&&\det\mathcal{C}_l^{(1,3(4))}=\mp t_{1,1}^{r}t_{2,N}^{i(r)}\chi_{2l-1,1}\chi_{2l,N},\nonumber\\
				&&\det\mathcal{C}_l^{(2,3(4))}=\mp t_{1,1}^{i}t_{2,N}^{i(r)}\chi_{2l-1,1}\chi_{2l,N}.\label{C1234_2C}
			\end{eqnarray}
			Thus, $\tilde{\mathcal{P}}_{2}(0) \neq 0$ for $N_{M} = 1$ and $N \to \infty$, i.e., unlike the single-lead case, the NESS remains non-degenerate.
			
			For $N_{M} = 2$, one obtains $\tilde{\mathcal{P}}_{2}(0) = 0$, since $\det\mathcal{C}_{1}^{(1,4)}  \det\mathcal{C}_{2}^{(2,3)} = \det\mathcal{C}_{1}^{(1,3)} \det\mathcal{C}_{2}^{(2,4)}$. To estimate the degeneracy of the NESS, we turn to the matrix $\tilde{\mathcal{B}}$,
			\begin{eqnarray}\label{Bdef_2C}
				&&\tilde{\mathcal{B}} = \left( \begin{array}{*{20}{c}}
					t_{1,1}^{r}\chi_{1,1} & t_{1,1}^{i}\chi_{1,1} & 0 & 0 \\
					0 & 0 & -t_{2,N}^{i}\chi_{2,N} & t_{2,N}^{r}\chi_{2,N} \\
					t_{1,1}^{r}\chi_{3,1} & t_{1,1}^{i}\chi_{3,1} & 0 & 0 \\
					0 & 0 & -t_{2,N}^{i}\chi_{4,N} & t_{2,N}^{r}\chi_{4,N}
				\end{array} \right).~~~
			\end{eqnarray}
			From the structure of the matrix in Eq. \eqref{Bdef_2C}, it follows that for $N_{M} \geq 2$ we have $\rk\tilde{\mathcal{B}} = 2$. According to Eq. \eqref{M0}, it results in $2(N_{M} - 1)$ zero-energy excitations in the open system.
			
			Consider the case 
			with a single lead interacting only with the first site of the chain and one bath affecting the entire system. In other words, in Eq. \eqref{l1234}, the second pair of fields is replaced by $\underline{\tilde{l}}_{3,4} = \underline{l}^{r,i}$. As a result, in contrast to the lead fields, the hybridization of the bath fields with the Majorana modes associated with $\varepsilon_{l} = 0$ generally yields $\mathcal{C}_l^{(3,4)} \neq 0$,
			\begin{eqnarray}\label{C1234_1C1B}
				&&\det\mathcal{C}_l^{(1,2)}=0,\nonumber\\
				&&\det\mathcal{C}_l^{(3,4)}=\left(l_{A,1}^{r}l_{B,N}^{i}-l_{A,1}^{i}l_{B,N}^{r}\right)\chi_{2l-1,1}\chi_{2l,N},\nonumber\\
				&&\det\mathcal{C}_l^{(1,3(4))}= t_{1,1}^{r}l_{B,N}^{r(i)}\chi_{2l-1,1}\chi_{2l,N},~\notag\\
				&&\det\mathcal{C}_l^{(2,3(4))}= t_{1,1}^{i}l_{B,N}^{r(i)}\chi_{2l-1,1}\chi_{2l,N}.
			\end{eqnarray}
			As a result, for $N_{M} = 1$, based on Eqs. \eqref{Pbet02} and \eqref{C1234_1C1B}, the NESS is non-degenerate.
			
			Next, for $N_{M} {=} 2$, degeneracy is achieved since $\tilde{\mathcal{P}}_{2}(0) = 0$ again due to the relation $\det\mathcal{C}_{1}^{(1,4)}\det\mathcal{C}_{2}^{(2,3)} = \det\mathcal{C}_{1}^{(1,3)} \det\mathcal{C}_{2}^{(2,4)}$. In this case, we find
			\begin{eqnarray}\label{Bdef_1C1B}
				&&\tilde{\mathcal{B}} = \left( \begin{array}{*{20}{c}}
					l_{A,1}^{r}\chi_{1,1} & l_{A,1}^{i}\chi_{1,1} & t_{1,1}^{r}\chi_{1,1} & t_{1,1}^{i}\chi_{1,1} \\
					l_{B,N}^{r}\chi_{2,N} & l_{B,N}^{i}\chi_{2,N} & 0 & 0 \\
					l_{A,1}^{r}\chi_{3,1} & l_{A,1}^{i}\chi_{3,1} & t_{1,1}^{r}\chi_{3,1} & t_{1,1}^{i}\chi_{3,1} \\
					l_{B,N}^{r}\chi_{4,N} & l_{B,N}^{i}\chi_{4,N} & 0 & 0
				\end{array} \right).~~~~~~
			\end{eqnarray}
			Therefore, the NESS degeneracy is equal to $2$ if no additional constraints are imposed on the bath fields. Generalization to the case of $N_{M} > 2$ and $N \to \infty$ results in $2N_{M} - 3$ zero-energy excitations in the open system.
			
			One scenario in which the degeneracy of the NESS increases further is the case of a Hermitian bath, $\hat{L}_{v} = \hat{L}_{v}^{+}$. As a result, $\underline{\nu} = \underline{\mu}^{*}$ and, according to the definition \eqref{lri_ferm0}, we have $\underline{l^{i}} = 0$. In this case, the rank of $\tilde{\mathcal{B}}$ decreases by $1$ (see Eq. \eqref{Bdef_1C1B}). The properties of $\tilde{\mathcal{B}}$ are analyzed in detail in Ref. \cite{shustin-25}.
			
			Summarizing the results presented in this Appendix, we provide two tables \ref{tab1} and \ref{tab2} corresponding to the cases $N_{B}+N_{L}=1$ and $N_{B}+N_{L}=2$, respectively. As in the discussion above, the distinction between the lead and bath lies in the fact that the former acts only on the edge site of the system, while the latter acts on the entire system. No additional conditions are imposed on the amplitudes of the fields. The presented results illustrate the applicability of the general expression \eqref{M}.
			
			\begin{table}[!htb]
				\begin{center}
					\caption{\label{tab1} The number of zero-energy excitations in open 1D superconducting system for $N_{B}+N_{L}=1$}
					\begin{tabular}{ |c|c|c| } 
						\hline
						& $N_{B}=1$ & $N_{L}=1$ \\
						\hline
						$N_{M}=0$ & $0$ & $0$ \\
						\hline
						$N_{M}\geq1$ & $2\left(N_{M}-1\right)$ & $2\left(N_{M}-1\right)+1$ \\
						\hline
					\end{tabular}
				\end{center}
			\end{table}
			
			\begin{table}[!htb]
				\begin{center}
					\caption{\label{tab2} The number of zero-energy excitations in open 1D superconducting system for $N_{B}+N_{L}=2$}
					\begin{tabular}{ |c|c|c|c| } 
						\hline
						& $N_{B}=2$ & $N_{L}=2$ & $N_{B}=1$, $N_{L}=1$ \\
						\hline
						$N_{M}\leq1$ & $0$ & $0$ & $0$ \\
						\hline
						$N_{M}>1$ & $2\left(N_{M}-2\right)$ & $2\left(N_{M}-2\right)+2$ & $2\left(N_{M}-2\right)+1$  \\
						\hline 
					\end{tabular}
				\end{center}
			\end{table}

			\bibliography{Majorana}
			
		\end{document}